\documentclass[twocolumn,preprintnumbers,amsmath,amssymb,superscriptaddress,prb,longbibliography]{revtex4-2}
\usepackage[dvipdfmx]{graphicx}
\usepackage{dcolumn}
\usepackage{bm}
\usepackage{color}
\usepackage{braket}

\begin{document}


\title{Topological surface states hybridized with bulk states of Bi-doped PbSb$_2$Te$_4$ revealed in quasiparticle interference
}

\author{Yuya Hattori}
\email{HATTORI.Yuya@nims.go.jp}
\affiliation{Research Center for Materials Nanoarchitectonics (MANA), National Institute for Materials Science, 3-13 Sakura, Tsukuba, Ibaraki 305-0003, Japan}
\affiliation{Institute of Industrial Science, The University of Tokyo, Komaba, Meguro-ku, Tokyo 153-8505, Japan}
\author{Keisuke Sagisaka}
\email{SAGISAKA.Keisuke@nims.go.jp}
\affiliation{Center for Basic Research on Materials, National Institute for Materials Science, 1-2-1 Sengen, Tsukuba, Ibaraki 305-0047, Japan.}
\author{Shunsuke Yoshizawa}
\affiliation{Center for Basic Research on Materials, National Institute for Materials Science, 1-2-1 Sengen, Tsukuba, Ibaraki 305-0047, Japan.}
\author{Yuki Tokumoto}
\affiliation{Institute of Industrial Science, The University of Tokyo, Komaba, Meguro-ku, Tokyo 153-8505, Japan}
\author{Keiichi Edagawa}
\affiliation{Institute of Industrial Science, The University of Tokyo, Komaba, Meguro-ku, Tokyo 153-8505, Japan}
\date{\today}

\begin{abstract}
Topological surface states of Bi-doped PbSb$_2$Te$_4$ [Pb(Bi$_{0.20}$Sb$_{0.80}$)$_{2}$Te$_{4}$] are investigated through analyses of quasiparticle interference (QPI) patterns observed by scanning tunneling microscopy. Interpretation of the experimental QPI patterns in the reciprocal space is achieved by numerical QPI simulations using two types of surface density of states produced by density functional theory calculations or the Fu's  surface state model [Phys. Rev. Lett. 103, 266801 (2009)]. We found that the Dirac point (DP) of the surface state appears in the bulk band gap of this material and, with the energy being away from the DP, the isoenergy contour of the surface state is substantially deformed or separated into segments due to hybridization with bulk electronic states. These findings provide a more accurate picture of topological surface states, especially at energies away from the DP, providing valuable insight into the electronic properties of topological insulators.
\end{abstract}


\maketitle

Topological insulators (TIs) with gapless surface states have attracted broad attention from both scientific and industrial fields. Soon after TIs were predicted \cite{Kane2005,Moore2007}, the presence of topological surface states was experimentally verified \cite{Hsieh2008}. Because of  the prohibition of backscattering in topological surface states with spin-polarized band dispersion \cite{Ando2013}, TIs are promising for use in high mobility spintronics devices. However, experimentally measured mobilities in topological surface states \cite{Xiong2012, Ren2010} are not as high as those in graphene with similar linear dispersions \cite{Tombros2011}. One of the reasons for the degraded mobility is hybridization between the topological surface states and bulk states \cite{Saha2014, Hsu2014}. A theoretical study has shown that topological surface states will lose their surface-localized nature because of hybridization with bulk states, resulting in an increase of the scattering rate \cite{Saha2014}. Nevertheless, the details of hybridization have not yet been experimentally verified, despite its importance for better understanding the scattering phenomena. In addition, the position of the Fermi energy ($E_F$) with respect to the Dirac point (DP) can also affect the mobility. Because the scattering of carriers at low temperatures is governed by defect scattering and the scattering rate is proportional to the density of states (DOS) \cite{Ando2009}, it is preferable for the $E_F$ to be located near the DP to achieve higher mobility. In many TIs, however, the DPs of the surface states are buried in bulk states \cite{Hsieh2008,Hsieh2009,Hsieh2009B,Chen2009,Xia2009, Neupane2012, Okuda2013}.

\begin{figure}
	\includegraphics[width=8.0cm]{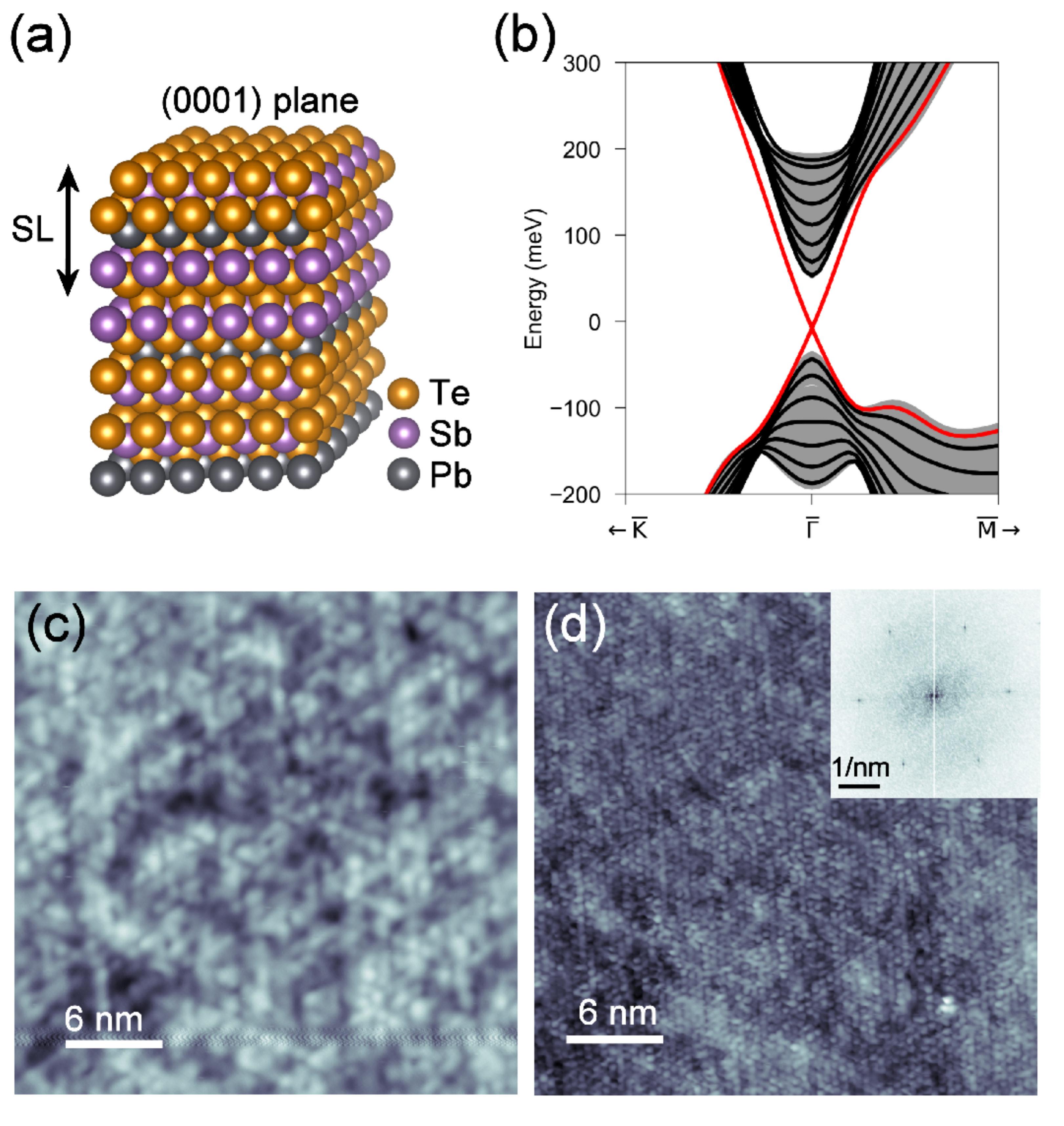}
	\caption{(a) The crystal structure of PbSb$_{2}$Te$_{4}$. A seven-layer unit (SL) is indicated by an arrow. (b) Band structure of PbSb$_{2}$Te$_{4}$ calculated with a slab model. Shaded regions represent bands calculated with a bulk cell. (c),(d) Constant current images of the (0001) surface of Bi-doped PbSb$_2$Te$_4$ in the identical region. The inset in (d) displays the FT of the constant-current image. Scan parameters: (c) $I$= 20 pA, $V_\textrm{s}$ = 500 mV, (d) $I$ = 1000 pA, $V_\textrm{s}$ = 50 mV. Image size: 29.3 $\times$ 29.3 nm$^2$}
	\label{fig1}
\end{figure}

PbSb$_2$Te$_4$ [Fig. \ref{fig1}(a)], which is known as a thermoelectric material \cite{Ikeda2007,Shelimova2007}, has been theoretically rediscovered as a TI with topological indices ($\nu_{0};\nu_{1} \nu_{2} \nu_{3}$)=(1;111) \cite{Jin2011, MENSHCHIKOVA2013}. In addition, this material has been predicted to bear its DP in the bulk band gap [Fig. \ref{fig1}(b)]\cite{MENSHCHIKOVA2013}, which is ideal for device applications. Souma \textit{et al}. conducted a survey of the band structure of Pb(Bi$_{1-x}$Sb$_x$)$_2$Te$_4$ using angle resolved photoemission spectroscopy (ARPES) \cite{Souma2012}. They reported that the $E_F$ of the alloy energetically shifts with respect to the DP as the mixing ratio between Bi and Sb in the alloy is varied. Consequently, the carrier type of the surface state changes from $n$-type for $x$ = 0 to $p$-type for $x$ = 1.0. These results demonstrate the potential of this alloy for realizing topological $pn$ junctions \cite{Wang2012}; however, isolation of the DP from the bulk states of PbSb$_2$Te$_4$ has not been experimentally verified because the unoccupied state is not accessible by ARPES. In addition to controlling the carrier type \cite{Hattori2017}, doping of Bi is also effective in stabilizing the crystal, because PbSb$_2$Te$_4$ is thermodynamically unstable \cite{Ikeda2007B}. 
 
In this Letter, we present the electronic character of surface states of Bi-doped PbSb$_2$Te$_4$ [Pb(Bi$_{0.20}$Sb$_{0.80}$)$_{2}$Te$_{4}$] clarified through analyses of quasiparticle interference (QPI) patterns recorded by scanning tunneling microscopy (STM). To interpret experimental QPI patterns, we conducted QPI pattern simulations using two types of surface DOS produced by density functional theory (DFT) calculations and the Fu's surface state model \cite{Fu2009}. As a result, we confirmed that the DP of the topological
surface state is well isolated from the bulk states of this material. Our DFT calculations show a prominent decrease in the surface DOS around the $\overline{\Gamma} \, \overline{M}$ direction in the energies where the surface states are overlapped by the bulk conduction bands (BCBs). As a consequence, QPI is limited by scattering between one side of the hexagonal Dirac cone in the $\overline{\Gamma} \, \overline{K}$ direction and its second-nearest sides in the corresponding energy range. The present study reveals that the Dirac cone of topological surface states at energies near the bulk states actually differs from the warped Dirac cone predicted by the Fu's surface state model \cite{Fu2009}. 
Methods for sample preparation and characterizations, as well as DFT calculations are described in the Supplemental Material (SM) \cite{Suppl}.

Figures 1(c) and 1(d) show constant-current images observed in the same area of a cleaved surface of Bi-doped PbSb$_2$Te$_4$. The image at a sample bias ($V_\textrm{s}$) of 500 mV shows an inhomogeneous contrast across the surface with slightly recognizable three-fold symmetry [Fig. \ref{fig1}(c)]. By contrast, an imaging condition with a smaller tip-surface distance ($V_\textrm{s}$ = 50 mV) achieved atomic resolution, showing each atom with different contrast [Fig. \ref{fig1}(d)]. This observation indicates that the cleaved surface (Te-terminated plane) is physically flat, with good single crystallinity. The crystallinity can be confirmed via the Fourier transform (FT) of the image, which revealed six sharp spots corresponding to the atomic lattice, without any additional features [inset of Fig. \ref{fig1}(d)]. However, the electronically disordered surface reflects the imperfection in the stoichiometry of the crystal due to randomly distributed antisite defects \cite{Note}. We attribute the electronically inhomogeneous contrast in the STM images to such antisite defects.

The electronic properties of the Bi-doped PbSb$_2$Te$_4$ surface were investigated using spectroscopic measurements. Figure 2(a) shows a d$I$/d$V$ spectrum of the sample surface (the variation of the spectral features across the surface is shown in the SM \cite{Suppl}). We found the DP of the surface state at a sample bias ($V_\textrm{s}$) of  +100 mV, which is likely to be completely isolated from the bulk states. By fitting linear functions to the spectrum \cite{Feenstra2005}, we estimated the top of the bulk valence band (BVB) and the bottom of the BCB to be located 77 $\pm$4 meV below and 86$\pm$13 meV above the DP, respectively. Accordingly, we derived a band gap of approximately 163 meV, which is comparatively smaller than that of PbBi$_2$Te$_4$ (200--230 meV) determined by ARPES \cite{Souma2012, Kuroda2012}. Nevertheless, this outcome provides further evidence for the existence of the DP within the bulk band gap of Bi-doped PbSb$_2$Te$_4$.

\begin{figure*}
	\includegraphics[width=17cm]{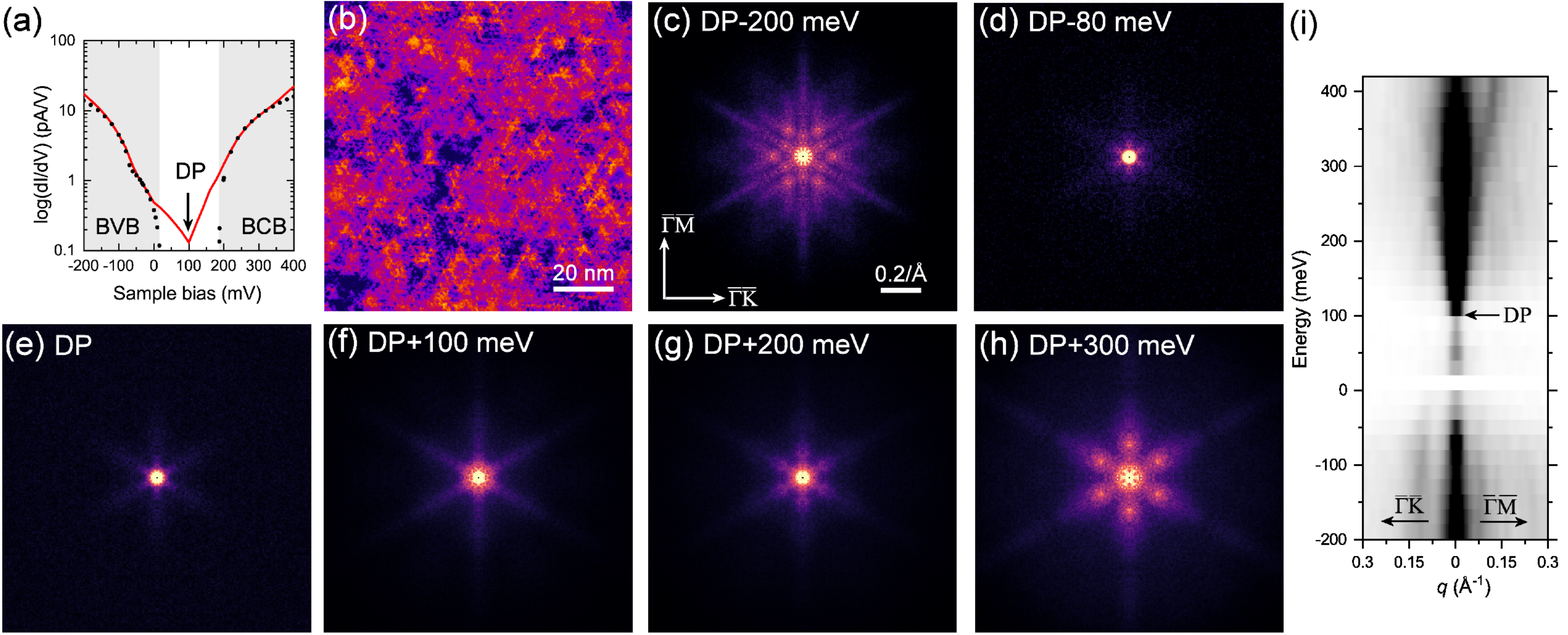}
	\caption{(a) Tunneling spectrum of Pb(Bi$_{0.20}$Sb$_{0.80}$)$_{2}$Te$_{4}$ in the logarithm scale. The spectrum was obtained by averaging 512 $\times$ 512 spectra recorded over a surface area of 100 $\times$ 100 nm$^2$. Set point: $V_\textrm{s}$ = +400 meV, $I$ = 500 pA. Dots are fitted to the spectrum using linear functions to estimate the bulk band edges. (b) d$I$/d$V$ map of the same surface acquired with $V_\textrm{s}$ = +300 mV and $I$ = 100 pA. (c)-(h) QPI patterns in the reciprocal space for different energies. The images were obtained by the FT of d$I$/d$V$ maps divided by the corresponding current ($I$) maps. Additionally, these images were drift-corrected and symmetrized \cite{Suppl}. Energies with respect to the DP are denoted in each image. The original $dI/dV$ and $I$ maps were recorded at $V_\textrm{s}$ values of (c) $-$100, (d) +20, (e) +100, (f) +200 (g) +300, and (h) +400 mV. (i)QPI dispersion obtained by plotting the FFT profiles along $\overline{\Gamma} \, \overline{K}$ and $\overline{\Gamma} \, \overline{M}$ directions.}
	\label{fig2}
\end{figure*}

Figure \ref{fig2}(b) shows a d$I$/d$V$ map of the same surface at $V_\textrm{s}$ = +300 mV, exhibiting complex patterns formed by QPI. As in other studies of TI surfaces \cite{Zhang2009,Okada2011,Stolyarov2021,Nurmamat2013}, the scattering property in the surface state can be evaluated by the FT of the QPI patterns to the reciprocal space.  Figures \ref{fig2}(c)--\ref{fig2}(h) show the FT of  d$I$/d$V$ maps for different sample biases generated from a data set of the spectroscopic measurement. In the energy range corresponding to the bulk band gap, the QPI patterns are characterized as a core structure with six streaky patterns extending along the $\overline{\Gamma} \, \overline{M}$ directions [Figs. \ref{fig2}(d)--\ref{fig2}(f)]. Such a QPI pattern stems from the isotropic Dirac cone, indicating that the DP of Bi-doped PbSb$_2$Te$_4$ is well isolated from the bulk bands. The dispersion relation depicted in Fig. \ref{fig2}(i), derived from the QPI patterns, also substantiates the presence of a linear dispersion near the DP.

As the energy moves away from the DP, the surface state overlaps with the bulk states and the QPI patterns gain additional features. At 300 meV above the DP, six intense spots emerge over the streak along the $\overline{\Gamma} \, \overline{M}$ directions [Fig. \ref{fig2}(h)]. Note that Bi$_2$Te$_3$ \cite{Zhang2009,Stolyarov2021} and Bi$_2$Te$_2$Se \cite{Nurmamat2013} TIs also exhibit a QPI feature along the $\overline{\Gamma} \, \overline{M}$ directions at energies well above the DP, but the shape of the feature was elongated in the $\overline{\Gamma} \, \overline{K}$ directions, unlike that observed in Fig. 2(h). At 200 meV below the DP, new spots with a V-shape pattern are observed in the ${\overline{\Gamma} \, \overline{K}}$ directions [Fig. \ref{fig2}(c)]. Such a new feature begins to arise at 100 meV below the DP, corresponding to the $E_F$ of the sample [Fig. S5(e) in the SM \cite{Suppl}]. We ascribe the emergence of these features that appear at both higher and lower energies away from the DP to hybridization of the surface state with the bulk states.

\begin{figure*}
	\includegraphics[width=15cm]{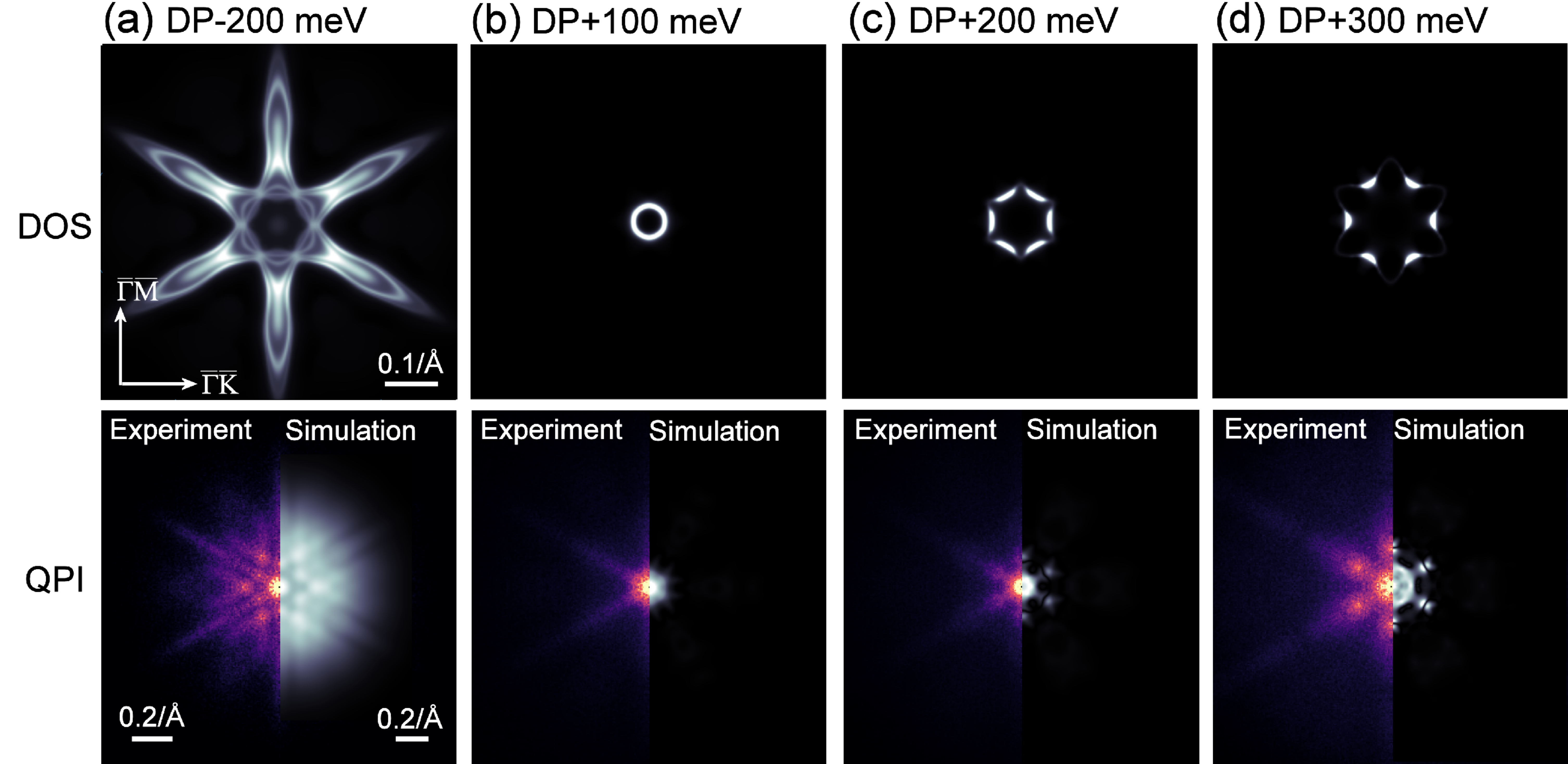}
	\caption{Isoenegy contour of the surface DOS and comparisons between experimental and simulated QPI. The surface DOS was produced from the result of the DFT calculations using a PbSb$_2$Te$_4$ slab. The surface DOS in (a) includes both the surface and bulk bands, whereas those in (b)--(d) include only the surface band. Characteristic experimental QPI patterns are selected from Fig. \ref{fig2}. The simulated QPI image in (a) is displayed after 25\% size reduction to attain good agreement with the experimental image. Other QPI images in (b)--(d) are displayed in the same scale between the experiment and simulation.}
	\label{fig3}
\end{figure*}

To clarify the formation process of the QPI patterns, we performed QPI simulations based on the $T$-matrix formalism \cite{Kohsaka2017} using spin-polarized eigenstates and eigenvalues obtained from DFT calculations. The surface DOS was computed in the reciprocal space as the product of the DOS and the Mulliken population of the Te atom on the surface. Further details regarding the simulation method can be found in the SM \cite{Suppl}. Isoenergy contour maps of the obtained surface DOS are displayed in the upper panel of Fig. \ref{fig3}. At 100 meV above the DP, the Dirac cone still exhibits a circular shape even though the BCB arises at this energy [Fig. \ref{fig3}(b)]. At 200 meV above the DP [Fig. \ref{fig3}(c)], the Dirac cone is deformed to a hexagonal shape by the warping effect \cite{Fu2009}. Note that the DOS at the hexagon vertices in the $\overline{\Gamma} \, \overline{M}$ lines is diminished. This tendency is enhanced at 300 meV above the DP [Fig. \ref{fig3}(d)], where more portions of the Dirac cone disappear in the vicinity of the $\overline{\Gamma} \, \overline{M}$ lines. Such a reduction of the surface DOS at the vertices of a warped Dirac cone has been suggested in the analyses of QPI patterns in Bi$_2$Te$_3$ \cite{Zhang2009,Wang2011} and in a DFT calculation for Bi$_2$Te$_2$Se \cite{Nurmamat2013}. We have observed that the surface state at $k$-points where the bulk bands overlap has undergone a reduction in localization, resulting in the dispersion of its charge density into the bulk (Fig. S6 in the SM \cite{Suppl}). This phenomenon is responsible for the partially missing DOS in the Dirac cone. By contrast, at 200 meV below the DP [Fig. \ref{fig3}(a)], the surface DOS is extremely distorted, with a shape that differs substantially from that of the Dirac cone. Actually, this surface DOS includes both the surface and bulk bands at this energy, whereas the other surface DOS in Figs. \ref{fig3}(b)--(d) include only the surface band at the corresponding energy [the red curves in Fig. \ref{fig1}(b)]. This treatment resulted in a good agreement between experimental and simulated QPI patterns. Consequently, this outcome suggests that the Dirac cone has a stronger tendency to hybridize with the BVBs than with the BCBs. These modifications of the surface state are attributed to hybridization with bulk states. When the bulk states are present, the Dirac cone completely disappears below the DP and is partially modified above the DP as the energy changes away from the DP. 

The lower panel in Fig. \ref{fig3} shows a comparison of Fourier transformed QPI patterns between the experiment and simulation. Our QPI simulation using the surface DOS produced by DFT calculations shows acceptable agreement with the experimental QPI patterns. The feature with the maxima appearing in the $\overline{\Gamma} \, \overline{M}$ directions at 200 and 300 meV above the DP are well reproduced in the simulation. The simulated QPI pattern at 200 meV below the DP also reproduced the general structures observed in the experiment, although the size of the overall QPI pattern needed to be reduced by 25\% to fit the experimental pattern \cite{Note1}. 
QPI patterns observed at energies further away from the DP are presented in Fig. S5 of the SM \cite{Suppl}.

\begin{figure}
	\includegraphics[width=8.5cm]{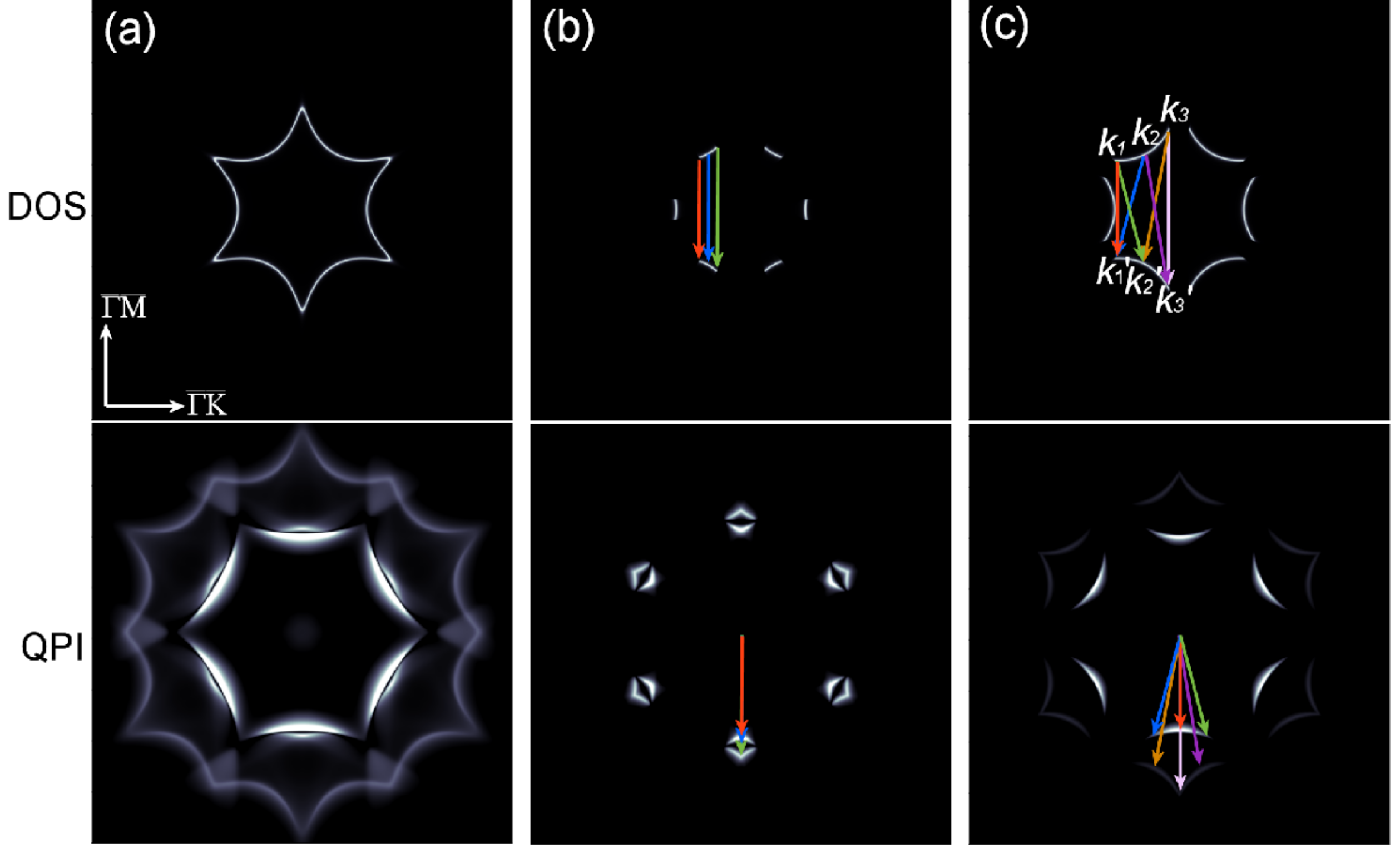}
	\caption{Isoenergy contour of surface DOS created by Fu's model \cite{Fu2009} and the corresponding QPI patterns in the reciprocal space at 300 meV above the DP. The surface DOS in (a) was computed by the Fu's Hamiltonian \cite{Fu2009} using parameters $m^*$ = 0.092 eV$^{-1}$\AA$^{-2}$, $v$ = 0.5 eV\AA, and $\lambda$ = 20 eV\AA$^3$. The modified surface state DOSs in (b) and (c) were created by truncating the corner parts of the hexagonal DOS around the $\overline{\Gamma} \, \overline{M}$ directions in (a). Arrows in (b) and (c) indicate possible scattering wave vectors.}
	\label{fig4}
\end{figure}

To gain further insight into the scattering property of the topological surface states through the formation of QPI patterns, we modeled a surface state expressed by Fu's Hamiltonian \cite{Fu2009}. In this attempt, we prepared a moderately warped Dirac cone [Fig. \ref{fig4}(a)], which is similar in shape to the Dirac cone for Bi$_2$Te$_3$ \cite{Chen2009}. The corresponding QPI pattern, being nearly a hexagon with satellite fringes, does not precisely reproduce any experimental QPI patterns \cite{Zhang2009,Okada2011,Stolyarov2021}.

To reproduce the PbSb$_2$Te$_4$ surface DOS that resulted from hybridization with the BCB, as observed in Fig. \ref{fig3}(d), the vertices were largely discarded from the concave hexagram of the Dirac cone in Fig. \ref{fig4}(a). The corresponding QPI pattern in the lower panel of Fig. \ref{fig4}(b) shows a scattering pattern in the $\overline{\Gamma} \, \overline{M}$ directions, similar to the pattern in Fig. \ref{fig3}(d). Thus, the formation of QPI patterns on the Bi-doped PbSb$_2$Te$_4$ surface at energies near the bottom of the BCB is achieved by scattering between one segment of the Dirac cone centered at the $\overline{\Gamma} \, \overline{K}$ direction and its next-nearest-neighbors, as denoted by arrows in Fig. \ref{fig4}(b). We created another surface DOS that retains the warping feature by limiting truncation around the vertices of the concave hexagram. The corresponding QPI pattern exhibits a feature centering in the $\overline{\Gamma} \, \overline{M}$ directions but elongating to the $\overline{\Gamma} \, \overline{K}$ directions [lower panel of Fig. \ref{fig4}(c)]. Such a QPI pattern is similar to those observed for Bi$_2$Te$_3$ \cite{Okada2011, Stolyarov2021} and Bi$_2$Te$_2$Se \cite{Nurmamat2013}. In these cases, scattering predominantly occurs between $k$-point pairs [$k_1$, $k_1'$], [$k_1$, $k_2'$] and [$k_2$, $k_1'$] shown in Fig. \ref{fig4}(c). These results of our QPI analyses underscore that the shape of the QPI pattern is dependent on the surface state segmentation induced by bulk bands. Therefore, the dominant scattering wave vector differs among TIs. Considering the dominant scattering process on the Bi-doped PbSb$_2$Te$_4$ surface, the QPI dispersion displayed in Fig. \ref{fig2}(i) can covert to the surface band dispersion using the relation $q$ = $\sqrt{3}k_{\overline{\Gamma} \, \overline{K}}$. The resulting surface band can be characterized as having an almost linear dispersion with a group velocity of 2.2 eV\AA {} at the DP (see SM \cite{Suppl}), whereas the shape for PbBi$_2$Te$_4$ appears to be non-linear with a velocity of 1.0 eV\AA {} \cite{Kuroda2012, Souma2012}.

The present results of our QPI analyses shed new light on an improved understanding of the surface physics of TIs. The physical properties of TIs are generally discussed on the basis of the Fu's surface-state model \cite{Li2014,Akzyanov2018}; however, we have shown that the topological surface states cannot be described simply by a warped Dirac cone at energies in the vicinity of bulk state. The picture of the Dirac cone separated into segments at energies away from the DP will facilitate understanding of a weakening of the Landau quantization near the BCB of TIs \cite{Okada2012}. Moreover, surface--bulk hybridization is a critical concept in discussions of the enhanced scattering rate \cite{Saha2014} and phase-coherent transport between surfaces and bulk \cite{Steinberg2011,Tian2014}. Finally, surface--bulk hybridization is also important in the design of spintronic devices that exploit the anomalously high charge--spin current conversion efficiency observed near the bulk band edge of TIs \cite{Kondou2016,Khang2018}.

\begin{acknowledgments}
Computational simulations were performed on the NIMS numerical simulator. STM measurements were supported by the NIMS microstructural characterization platform (NMCP) as a program of "Nanotechnology Platform" of the Ministry of Education, Culture, Sports, Science and Technology (MEXT), Japan, Grant Number A-20-NM-0112. YH acknowledges financial support by JSPS KAKENHI (Grant Nos. 19J13968, 18H01692, and 19K04984), KS by the Murata Science Foundation, and SY by JSPS KAKENHI (Grant Nos. 20H05277, 21H01817). MANA was established by World Premier International Research Center Initiative (WPI), MEXT, Japan.
\end{acknowledgments}



\begin{thebibliography}{47}%
	\makeatletter
	\providecommand \@ifxundefined [1]{%
		\@ifx{#1\undefined}
	}%
	\providecommand \@ifnum [1]{%
		\ifnum #1\expandafter \@firstoftwo
		\else \expandafter \@secondoftwo
		\fi
	}%
	\providecommand \@ifx [1]{%
		\ifx #1\expandafter \@firstoftwo
		\else \expandafter \@secondoftwo
		\fi
	}%
	\providecommand \natexlab [1]{#1}%
	\providecommand \enquote  [1]{``#1''}%
	\providecommand \bibnamefont  [1]{#1}%
	\providecommand \bibfnamefont [1]{#1}%
	\providecommand \citenamefont [1]{#1}%
	\providecommand \href@noop [0]{\@secondoftwo}%
	\providecommand \href [0]{\begingroup \@sanitize@url \@href}%
	\providecommand \@href[1]{\@@startlink{#1}\@@href}%
	\providecommand \@@href[1]{\endgroup#1\@@endlink}%
	\providecommand \@sanitize@url [0]{\catcode `\\12\catcode `\$12\catcode
		`\&12\catcode `\#12\catcode `\^12\catcode `\_12\catcode `\%12\relax}%
	\providecommand \@@startlink[1]{}%
	\providecommand \@@endlink[0]{}%
	\providecommand \url  [0]{\begingroup\@sanitize@url \@url }%
	\providecommand \@url [1]{\endgroup\@href {#1}{\urlprefix }}%
	\providecommand \urlprefix  [0]{URL }%
	\providecommand \Eprint [0]{\href }%
	\providecommand \doibase [0]{https://doi.org/}%
	\providecommand \selectlanguage [0]{\@gobble}%
	\providecommand \bibinfo  [0]{\@secondoftwo}%
	\providecommand \bibfield  [0]{\@secondoftwo}%
	\providecommand \translation [1]{[#1]}%
	\providecommand \BibitemOpen [0]{}%
	\providecommand \bibitemStop [0]{}%
	\providecommand \bibitemNoStop [0]{.\EOS\space}%
	\providecommand \EOS [0]{\spacefactor3000\relax}%
	\providecommand \BibitemShut  [1]{\csname bibitem#1\endcsname}%
	\let\auto@bib@innerbib\@empty
	\bibitem [{\citenamefont {Kane}\ and\ \citenamefont {Mele}(2005)}]{Kane2005}%
	\BibitemOpen
	\bibfield  {author} {\bibinfo {author} {\bibfnamefont {C.~L.}\ \bibnamefont
			{Kane}}\ and\ \bibinfo {author} {\bibfnamefont {E.~J.}\ \bibnamefont
			{Mele}},\ }\bibfield  {title} {\bibinfo {title} {Quantum spin Hall effect in
			graphene},\ }\href@noop {} {\bibfield  {journal} {\bibinfo  {journal} {Phys.
				Rev. Lett.}\ }\textbf {\bibinfo {volume} {95}},\ \bibinfo {pages} {226801}
		(\bibinfo {year} {2005})}\BibitemShut {NoStop}%
	\bibitem [{\citenamefont {Moore}\ and\ \citenamefont
		{Balents}(2007)}]{Moore2007}%
	\BibitemOpen
	\bibfield  {author} {\bibinfo {author} {\bibfnamefont {J.~E.}\ \bibnamefont
			{Moore}}\ and\ \bibinfo {author} {\bibfnamefont {L.}~\bibnamefont
			{Balents}},\ }\bibfield  {title} {\bibinfo {title} {Topological invariants of
			time-reversal-invariant band structures},\ }\href@noop {} {\bibfield
		{journal} {\bibinfo  {journal} {Phys. Rev. B}\ }\textbf {\bibinfo {volume}
			{75}},\ \bibinfo {pages} {121306(R)} (\bibinfo {year} {2007})}\BibitemShut
	{NoStop}%
	\bibitem [{\citenamefont {Hsieh}\ \emph {et~al.}(2008)\citenamefont {Hsieh},
		\citenamefont {Qian}, \citenamefont {Wray}, \citenamefont {Xia},
		\citenamefont {Hor}, \citenamefont {Cava},\ and\ \citenamefont
		{Hasan}}]{Hsieh2008}%
	\BibitemOpen
	\bibfield  {author} {\bibinfo {author} {\bibfnamefont {D.}~\bibnamefont
			{Hsieh}}, \bibinfo {author} {\bibfnamefont {D.}~\bibnamefont {Qian}},
		\bibinfo {author} {\bibfnamefont {L.}~\bibnamefont {Wray}}, \bibinfo {author}
		{\bibfnamefont {Y.}~\bibnamefont {Xia}}, \bibinfo {author} {\bibfnamefont
			{Y.~S.}\ \bibnamefont {Hor}}, \bibinfo {author} {\bibfnamefont {R.~J.}\
			\bibnamefont {Cava}},\ and\ \bibinfo {author} {\bibnamefont {Hasan}},\
	}\bibfield  {title} {\bibinfo {title} {A topological Dirac insulator in a
			quantum spin Hall phase},\ }\href@noop {} {\bibfield  {journal} {\bibinfo
			{journal} {Nature}\ }\textbf {\bibinfo {volume} {452}},\ \bibinfo {pages}
		{970} (\bibinfo {year} {2008})}\BibitemShut {NoStop}%
	\bibitem [{\citenamefont {Ando}(2013)}]{Ando2013}%
	\BibitemOpen
	\bibfield  {author} {\bibinfo {author} {\bibfnamefont {Y.}~\bibnamefont
			{Ando}},\ }\bibfield  {title} {\bibinfo {title} {Topological insulator
			materials},\ }\href@noop {} {\bibfield  {journal} {\bibinfo  {journal}
			{J. Phys. Soc. Jpn.}\ }\textbf {\bibinfo {volume}
			{82}},\ \bibinfo {pages} {102001} (\bibinfo {year} {2013})}\BibitemShut
	{NoStop}%
	\bibitem [{\citenamefont {Xiong}\ \emph {et~al.}(2012)\citenamefont {Xiong},
		\citenamefont {Petersen}, \citenamefont {Qu}, \citenamefont {Horb},
		\citenamefont {Cava},\ and\ \citenamefont {Ong}}]{Xiong2012}%
	\BibitemOpen
	\bibfield  {author} {\bibinfo {author} {\bibfnamefont {J.}~\bibnamefont
			{Xiong}}, \bibinfo {author} {\bibfnamefont {A.~C.}\ \bibnamefont {Petersen}},
		\bibinfo {author} {\bibfnamefont {D.}~\bibnamefont {Qu}}, \bibinfo {author}
		{\bibfnamefont {Y.~S.}\ \bibnamefont {Horb}}, \bibinfo {author}
		{\bibfnamefont {R.~J.}\ \bibnamefont {Cava}},\ and\ \bibinfo {author}
		{\bibfnamefont {N.~P.}\ \bibnamefont {Ong}},\ }\bibfield  {title} {\bibinfo
		{title} {Quantum oscillations in a topological insulator Bi$_2$Te$_2$Se with large
			bulk resistivity (6 $\Omega$cm)},\ }\href@noop {} {\bibfield  {journal}
		{\bibinfo  {journal} {Physica E}\ }\textbf {\bibinfo {volume} {44}},\
		\bibinfo {pages} {917} (\bibinfo {year} {2012})}\BibitemShut {NoStop}%
	\bibitem [{\citenamefont {Ren}\ \emph {et~al.}(2010)\citenamefont {Ren},
		\citenamefont {Taskin}, \citenamefont {Sasaki}, \citenamefont {Segawa},\ and\
		\citenamefont {Ando}}]{Ren2010}%
	\BibitemOpen
	\bibfield  {author} {\bibinfo {author} {\bibfnamefont {Z.}~\bibnamefont
			{Ren}}, \bibinfo {author} {\bibfnamefont {A.~A.}\ \bibnamefont {Taskin}},
		\bibinfo {author} {\bibfnamefont {S.}~\bibnamefont {Sasaki}}, \bibinfo
		{author} {\bibfnamefont {K.}~\bibnamefont {Segawa}},\ and\ \bibinfo {author}
		{\bibfnamefont {Y.}~\bibnamefont {Ando}},\ }\bibfield  {title} {\bibinfo
		{title} {Large bulk resistivity and surface quantum oscillations in the
			topological insulator ${\text{Bi}}_{2}{\text{Te}}_{2}\text{Se}$},\
	}\href@noop {} {\bibfield  {journal} {\bibinfo  {journal} {Phys. Rev. B}\
		}\textbf {\bibinfo {volume} {82}},\ \bibinfo {pages} {241306(R)} (\bibinfo
		{year} {2010})}\BibitemShut {NoStop}%
	\bibitem [{\citenamefont {Tombros}\ \emph {et~al.}(2011)\citenamefont
		{Tombros}, \citenamefont {Veligura}, \citenamefont {Junesch}, \citenamefont
		{Jasper van~den Berg}, \citenamefont {Zomer}, \citenamefont {Wojtaszek},
		\citenamefont {Vera~Marun}, \citenamefont {Jonkman},\ and\ \citenamefont {van
			Wees}}]{Tombros2011}%
	\BibitemOpen
	\bibfield  {author} {\bibinfo {author} {\bibfnamefont {N.}~\bibnamefont
			{Tombros}}, \bibinfo {author} {\bibfnamefont {A.}~\bibnamefont {Veligura}},
		\bibinfo {author} {\bibfnamefont {J.}~\bibnamefont {Junesch}}, \bibinfo
		{author} {\bibfnamefont {J.}~\bibnamefont {Jasper van~den Berg}}, \bibinfo
		{author} {\bibfnamefont {P.~J.}\ \bibnamefont {Zomer}}, \bibinfo {author}
		{\bibfnamefont {M.}~\bibnamefont {Wojtaszek}}, \bibinfo {author}
		{\bibfnamefont {I.~J.}\ \bibnamefont {Vera~Marun}}, \bibinfo {author}
		{\bibfnamefont {H.~T.}\ \bibnamefont {Jonkman}},\ and\ \bibinfo {author}
		{\bibfnamefont {B.~J.}\ \bibnamefont {van Wees}},\ }\bibfield  {title}
	{\bibinfo {title} {Large yield production of high mobility freely suspended
			graphene electronic devices on a polydimethylglutarimide based organic
			polymer},\ }\href@noop {} {\bibfield  {journal} {\bibinfo  {journal} {J. Appl. Phys.}\ }\textbf {\bibinfo {volume} {109}},\ \bibinfo {pages}
		{093702} (\bibinfo {year} {2011})}\BibitemShut {NoStop}%
	\bibitem [{\citenamefont {Saha}\ and\ \citenamefont {Garate}(2014)}]{Saha2014}%
	\BibitemOpen
	\bibfield  {author} {\bibinfo {author} {\bibfnamefont {K.}~\bibnamefont
			{Saha}}\ and\ \bibinfo {author} {\bibfnamefont {I.}~\bibnamefont {Garate}},\
	}\bibfield  {title} {\bibinfo {title} {Theory of bulk-surface coupling in
			topological insulator films},\ }\href@noop {} {\bibfield  {journal} {\bibinfo
			{journal} {Phys. Rev. B}\ }\textbf {\bibinfo {volume} {90}},\ \bibinfo
		{pages} {245418} (\bibinfo {year} {2014})}\BibitemShut {NoStop}%
	\bibitem [{\citenamefont {Hsu}\ \emph {et~al.}(2014)\citenamefont {Hsu},
		\citenamefont {Fischer}, \citenamefont {Hughes}, \citenamefont {Park},\ and\
		\citenamefont {Kim}}]{Hsu2014}%
	\BibitemOpen
	\bibfield  {author} {\bibinfo {author} {\bibfnamefont {Y.-T.}\ \bibnamefont
			{Hsu}}, \bibinfo {author} {\bibfnamefont {M.~H.}\ \bibnamefont {Fischer}},
		\bibinfo {author} {\bibfnamefont {T.~L.}\ \bibnamefont {Hughes}}, \bibinfo
		{author} {\bibfnamefont {K.}~\bibnamefont {Park}},\ and\ \bibinfo {author}
		{\bibfnamefont {E.-A.}\ \bibnamefont {Kim}},\ }\bibfield  {title} {\bibinfo
		{title} {Effects of surface-bulk hybridization in three-dimensional
			topological metals},\ }\href@noop {} {\bibfield  {journal} {\bibinfo
			{journal} {Phys. Rev. B}\ }\textbf {\bibinfo {volume} {89}},\ \bibinfo
		{pages} {205438} (\bibinfo {year} {2014})}\BibitemShut {NoStop}%
	\bibitem [{\citenamefont {Ando}(2009)}]{Ando2009}%
	\BibitemOpen
	\bibfield  {author} {\bibinfo {author} {\bibfnamefont {T.}~\bibnamefont
			{Ando}},\ }\bibfield  {title} {\bibinfo {title} {The electronic properties of
			graphene and carbon nanotubes},\ }\href@noop {} {\bibfield  {journal}
		{\bibinfo  {journal} {NPG Asia Materials}\ }\textbf {\bibinfo {volume} {1}},\
		\bibinfo {pages} {17} (\bibinfo {year} {2009})}\BibitemShut {NoStop}%
	\bibitem [{\citenamefont {Hsieh}\ \emph
		{et~al.}(2009{\natexlab{a}})\citenamefont {Hsieh}, \citenamefont {Xia},
		\citenamefont {Qian}, \citenamefont {Wray}, \citenamefont {Dil},
		\citenamefont {Meier}, \citenamefont {Osterwalder}, \citenamefont {Patthey},
		\citenamefont {Checkelsky}, \citenamefont {Ong}, \citenamefont {Fedorov},
		\citenamefont {Lin}, \citenamefont {Bansil}, \citenamefont {Grauer},
		\citenamefont {Hor}, \citenamefont {Cava},\ and\ \citenamefont
		{Hasan}}]{Hsieh2009}%
	\BibitemOpen
	\bibfield  {author} {\bibinfo {author} {\bibfnamefont {D.}~\bibnamefont
			{Hsieh}}, \bibinfo {author} {\bibfnamefont {Y.}~\bibnamefont {Xia}}, \bibinfo
		{author} {\bibfnamefont {D.}~\bibnamefont {Qian}}, \bibinfo {author}
		{\bibfnamefont {L.}~\bibnamefont {Wray}}, \bibinfo {author} {\bibfnamefont
			{J.~H.}\ \bibnamefont {Dil}}, \bibinfo {author} {\bibfnamefont
			{F.}~\bibnamefont {Meier}}, \bibinfo {author} {\bibfnamefont
			{J.}~\bibnamefont {Osterwalder}}, \bibinfo {author} {\bibfnamefont
			{L.}~\bibnamefont {Patthey}}, \bibinfo {author} {\bibfnamefont {J.~G.}\
			\bibnamefont {Checkelsky}}, \bibinfo {author} {\bibfnamefont {N.~P.}\
			\bibnamefont {Ong}}, \bibinfo {author} {\bibfnamefont {A.~V.}\ \bibnamefont
			{Fedorov}}, \bibinfo {author} {\bibfnamefont {H.}~\bibnamefont {Lin}},
		\bibinfo {author} {\bibfnamefont {A.}~\bibnamefont {Bansil}}, \bibinfo
		{author} {\bibfnamefont {D.}~\bibnamefont {Grauer}}, \bibinfo {author}
		{\bibfnamefont {Y.~S.}\ \bibnamefont {Hor}}, \bibinfo {author} {\bibfnamefont
			{R.~J.}\ \bibnamefont {Cava}},\ and\ \bibinfo {author} {\bibfnamefont
			{M.~Z.}\ \bibnamefont {Hasan}},\ }\bibfield  {title} {\bibinfo {title} {A
			tunable topological insulator in the spin helical Dirac transport regime},\
	}\href@noop {} {\bibfield  {journal} {\bibinfo  {journal} {Nature}\ }\textbf
		{\bibinfo {volume} {460}},\ \bibinfo {pages} {1101} (\bibinfo {year}
		{2009}{\natexlab{a}})}\BibitemShut {NoStop}%
	\bibitem [{\citenamefont {Hsieh}\ \emph
		{et~al.}(2009{\natexlab{b}})\citenamefont {Hsieh}, \citenamefont {Xia},
		\citenamefont {Qian}, \citenamefont {Wray}, \citenamefont {Meier},
		\citenamefont {Dil}, \citenamefont {Osterwalder}, \citenamefont {Patthey},
		\citenamefont {Fedorov}, \citenamefont {Lin}, \citenamefont {Bansil},
		\citenamefont {Grauer}, \citenamefont {Hor}, \citenamefont {Cava},\ and\
		\citenamefont {Hasan}}]{Hsieh2009B}%
	\BibitemOpen
	\bibfield  {author} {\bibinfo {author} {\bibfnamefont {D.}~\bibnamefont
			{Hsieh}}, \bibinfo {author} {\bibfnamefont {Y.}~\bibnamefont {Xia}}, \bibinfo
		{author} {\bibfnamefont {D.}~\bibnamefont {Qian}}, \bibinfo {author}
		{\bibfnamefont {L.}~\bibnamefont {Wray}}, \bibinfo {author} {\bibfnamefont
			{F.}~\bibnamefont {Meier}}, \bibinfo {author} {\bibfnamefont {J.~H.}\
			\bibnamefont {Dil}}, \bibinfo {author} {\bibfnamefont {J.}~\bibnamefont
			{Osterwalder}}, \bibinfo {author} {\bibfnamefont {L.}~\bibnamefont
			{Patthey}}, \bibinfo {author} {\bibfnamefont {A.~V.}\ \bibnamefont
			{Fedorov}}, \bibinfo {author} {\bibfnamefont {H.}~\bibnamefont {Lin}},
		\bibinfo {author} {\bibfnamefont {A.}~\bibnamefont {Bansil}}, \bibinfo
		{author} {\bibfnamefont {D.}~\bibnamefont {Grauer}}, \bibinfo {author}
		{\bibfnamefont {Y.~S.}\ \bibnamefont {Hor}}, \bibinfo {author} {\bibfnamefont
			{R.~J.}\ \bibnamefont {Cava}},\ and\ \bibinfo {author} {\bibfnamefont
			{M.~Z.}\ \bibnamefont {Hasan}},\ }\bibfield  {title} {\bibinfo {title}
		{Observation of time-reversal-protected single-Dirac-cone
			topological-insulator states in ${\mathrm{Bi}}_{2}{\mathrm{Te}}_{3}$ and
			${\mathrm{Sb}}_{2}{\mathrm{Te}}_{3}$},\ }\href@noop {} {\bibfield  {journal}
		{\bibinfo  {journal} {Phys. Rev. Lett.}\ }\textbf {\bibinfo {volume} {103}},\
		\bibinfo {pages} {146401} (\bibinfo {year} {2009}{\natexlab{b}})}\BibitemShut
	{NoStop}%
	\bibitem [{\citenamefont {Chen}\ \emph {et~al.}(2009)\citenamefont {Chen},
		\citenamefont {Analytis}, \citenamefont {Chu}, \citenamefont {Liu},
		\citenamefont {Mo}, \citenamefont {Qi}, \citenamefont {Zhang}, \citenamefont
		{Lu}, \citenamefont {Dai}, \citenamefont {Fang}, \citenamefont {Zhang},
		\citenamefont {Fisher}, \citenamefont {Hussain},\ and\ \citenamefont
		{Shen}}]{Chen2009}%
	\BibitemOpen
	\bibfield  {author} {\bibinfo {author} {\bibfnamefont {Y.~L.}\ \bibnamefont
			{Chen}}, \bibinfo {author} {\bibfnamefont {J.~G.}\ \bibnamefont {Analytis}},
		\bibinfo {author} {\bibfnamefont {J.-H.}\ \bibnamefont {Chu}}, \bibinfo
		{author} {\bibfnamefont {Z.~K.}\ \bibnamefont {Liu}}, \bibinfo {author}
		{\bibfnamefont {S.-K.}\ \bibnamefont {Mo}}, \bibinfo {author} {\bibfnamefont
			{X.~L.}\ \bibnamefont {Qi}}, \bibinfo {author} {\bibfnamefont {H.~J.}\
			\bibnamefont {Zhang}}, \bibinfo {author} {\bibfnamefont {D.~H.}\ \bibnamefont
			{Lu}}, \bibinfo {author} {\bibfnamefont {X.}~\bibnamefont {Dai}}, \bibinfo
		{author} {\bibfnamefont {Z.}~\bibnamefont {Fang}}, \bibinfo {author}
		{\bibfnamefont {S.~C.}\ \bibnamefont {Zhang}}, \bibinfo {author}
		{\bibfnamefont {I.~R.}\ \bibnamefont {Fisher}}, \bibinfo {author}
		{\bibfnamefont {Z.}~\bibnamefont {Hussain}},\ and\ \bibinfo {author}
		{\bibfnamefont {Z.-X.}\ \bibnamefont {Shen}},\ }\bibfield  {title} {\bibinfo
		{title} {Experimental realization of a three-dimensional topological
			insulator, Bi$_2$Te$_3$},\ }\href@noop {}
	{\bibfield  {journal} {\bibinfo  {journal} {Science}\ }\textbf {\bibinfo
			{volume} {325}},\ \bibinfo {pages} {178} (\bibinfo {year}
		{2009})}\BibitemShut {NoStop}%
	\bibitem [{\citenamefont {Xia}\ \emph {et~al.}(2009)\citenamefont {Xia},
		\citenamefont {Qian}, \citenamefont {Hsieh}, \citenamefont {Wray},
		\citenamefont {Pal}, \citenamefont {Lin}, \citenamefont {Bansil},
		\citenamefont {Grauer}, \citenamefont {Hor}, \citenamefont {Cava},\ and\
		\citenamefont {Hasan}}]{Xia2009}%
	\BibitemOpen
	\bibfield  {author} {\bibinfo {author} {\bibfnamefont {Y.}~\bibnamefont
			{Xia}}, \bibinfo {author} {\bibfnamefont {D.}~\bibnamefont {Qian}}, \bibinfo
		{author} {\bibfnamefont {D.}~\bibnamefont {Hsieh}}, \bibinfo {author}
		{\bibfnamefont {L.}~\bibnamefont {Wray}}, \bibinfo {author} {\bibfnamefont
			{A.}~\bibnamefont {Pal}}, \bibinfo {author} {\bibfnamefont {H.}~\bibnamefont
			{Lin}}, \bibinfo {author} {\bibfnamefont {A.}~\bibnamefont {Bansil}},
		\bibinfo {author} {\bibfnamefont {D.}~\bibnamefont {Grauer}}, \bibinfo
		{author} {\bibfnamefont {Y.~S.}\ \bibnamefont {Hor}}, \bibinfo {author}
		{\bibfnamefont {R.~J.}\ \bibnamefont {Cava}},\ and\ \bibinfo {author}
		{\bibfnamefont {M.~Z.}\ \bibnamefont {Hasan}},\ }\bibfield  {title} {\bibinfo
		{title} {Observation of a large-gap topological-insulator class with a single
			dirac cone on the surface},\ }\href@noop {} {\bibfield  {journal} {\bibinfo
			{journal} {Nature Physics}\ }\textbf {\bibinfo {volume} {5}},\ \bibinfo
		{pages} {398} (\bibinfo {year} {2009})}\BibitemShut {NoStop}%
	\bibitem [{\citenamefont {Neupane}\ \emph {et~al.}(2012)\citenamefont
		{Neupane}, \citenamefont {Xu}, \citenamefont {Wray}, \citenamefont
		{Petersen}, \citenamefont {Shankar}, \citenamefont {Alidoust}, \citenamefont
		{Liu}, \citenamefont {Fedorov}, \citenamefont {Ji}, \citenamefont {Allred},
		\citenamefont {Hor}, \citenamefont {Chang}, \citenamefont {Jeng},
		\citenamefont {Lin}, \citenamefont {Bansil}, \citenamefont {Cava},\ and\
		\citenamefont {Hasan}}]{Neupane2012}%
	\BibitemOpen
	\bibfield  {author} {\bibinfo {author} {\bibfnamefont {M.}~\bibnamefont
			{Neupane}}, \bibinfo {author} {\bibfnamefont {S.-Y.}\ \bibnamefont {Xu}},
		\bibinfo {author} {\bibfnamefont {L.~A.}\ \bibnamefont {Wray}}, \bibinfo
		{author} {\bibfnamefont {A.}~\bibnamefont {Petersen}}, \bibinfo {author}
		{\bibfnamefont {R.}~\bibnamefont {Shankar}}, \bibinfo {author} {\bibfnamefont
			{N.}~\bibnamefont {Alidoust}}, \bibinfo {author} {\bibfnamefont
			{C.}~\bibnamefont {Liu}}, \bibinfo {author} {\bibfnamefont {A.}~\bibnamefont
			{Fedorov}}, \bibinfo {author} {\bibfnamefont {H.}~\bibnamefont {Ji}},
		\bibinfo {author} {\bibfnamefont {J.~M.}\ \bibnamefont {Allred}}, \bibinfo
		{author} {\bibfnamefont {Y.~S.}\ \bibnamefont {Hor}}, \bibinfo {author}
		{\bibfnamefont {T.-R.}\ \bibnamefont {Chang}}, \bibinfo {author}
		{\bibfnamefont {H.-T.}\ \bibnamefont {Jeng}}, \bibinfo {author}
		{\bibfnamefont {H.}~\bibnamefont {Lin}}, \bibinfo {author} {\bibfnamefont
			{A.}~\bibnamefont {Bansil}}, \bibinfo {author} {\bibfnamefont {R.~J.}\
			\bibnamefont {Cava}},\ and\ \bibinfo {author} {\bibfnamefont {M.~Z.}\
			\bibnamefont {Hasan}},\ }\bibfield  {title} {\bibinfo {title} {Topological
			surface states and Dirac point tuning in ternary topological insulators},\
	}\href@noop {} {\bibfield  {journal} {\bibinfo  {journal} {Phys. Rev. B}\
		}\textbf {\bibinfo {volume} {85}},\ \bibinfo {pages} {235406} (\bibinfo
		{year} {2012})}\BibitemShut {NoStop}%
	\bibitem [{\citenamefont {Okuda}\ \emph {et~al.}(2013)\citenamefont {Okuda},
		\citenamefont {Maegawa}, \citenamefont {Ye}, \citenamefont {Shirai},
		\citenamefont {Warashina}, \citenamefont {Miyamoto}, \citenamefont {Kuroda},
		\citenamefont {Arita}, \citenamefont {Aliev}, \citenamefont {Amiraslanov},
		\citenamefont {Babanly}, \citenamefont {Chulkov}, \citenamefont {Eremeev},
		\citenamefont {Kimura}, \citenamefont {Namatame},\ and\ \citenamefont
		{Taniguchi}}]{Okuda2013}%
	\BibitemOpen
	\bibfield  {author} {\bibinfo {author} {\bibfnamefont {T.}~\bibnamefont
			{Okuda}}, \bibinfo {author} {\bibfnamefont {T.}~\bibnamefont {Maegawa}},
		\bibinfo {author} {\bibfnamefont {M.}~\bibnamefont {Ye}}, \bibinfo {author}
		{\bibfnamefont {K.}~\bibnamefont {Shirai}}, \bibinfo {author} {\bibfnamefont
			{T.}~\bibnamefont {Warashina}}, \bibinfo {author} {\bibfnamefont
			{K.}~\bibnamefont {Miyamoto}}, \bibinfo {author} {\bibfnamefont
			{K.}~\bibnamefont {Kuroda}}, \bibinfo {author} {\bibfnamefont
			{M.}~\bibnamefont {Arita}}, \bibinfo {author} {\bibfnamefont {Z.~S.}\
			\bibnamefont {Aliev}}, \bibinfo {author} {\bibfnamefont {I.~R.}\ \bibnamefont
			{Amiraslanov}}, \bibinfo {author} {\bibfnamefont {M.~B.}\ \bibnamefont
			{Babanly}}, \bibinfo {author} {\bibfnamefont {E.~V.}\ \bibnamefont
			{Chulkov}}, \bibinfo {author} {\bibfnamefont {S.~V.}\ \bibnamefont
			{Eremeev}}, \bibinfo {author} {\bibfnamefont {A.}~\bibnamefont {Kimura}},
		\bibinfo {author} {\bibfnamefont {H.}~\bibnamefont {Namatame}},\ and\
		\bibinfo {author} {\bibfnamefont {M.}~\bibnamefont {Taniguchi}},\ }\bibfield
	{title} {\bibinfo {title} {Experimental evidence of hidden topological
			surface states in ${\mathrm{PbBi}}_{4}{\mathrm{Te}}_{7}$},\ }\href@noop {}
	{\bibfield  {journal} {\bibinfo  {journal} {Phys. Rev. Lett.}\ }\textbf
		{\bibinfo {volume} {111}},\ \bibinfo {pages} {206803} (\bibinfo {year}
		{2013})}\BibitemShut {NoStop}%
	\bibitem [{\citenamefont {Ikeda}\ \emph
		{et~al.}(2007{\natexlab{a}})\citenamefont {Ikeda}, \citenamefont {Haile},
		\citenamefont {Ravi}, \citenamefont {Azizgolshani}, \citenamefont {Gascoin},\
		and\ \citenamefont {Snyder}}]{Ikeda2007}%
	\BibitemOpen
	\bibfield  {author} {\bibinfo {author} {\bibfnamefont {T.}~\bibnamefont
			{Ikeda}}, \bibinfo {author} {\bibfnamefont {S.~M.}\ \bibnamefont {Haile}},
		\bibinfo {author} {\bibfnamefont {V.~A.}\ \bibnamefont {Ravi}}, \bibinfo
		{author} {\bibfnamefont {H.}~\bibnamefont {Azizgolshani}}, \bibinfo {author}
		{\bibfnamefont {F.}~\bibnamefont {Gascoin}},\ and\ \bibinfo {author}
		{\bibfnamefont {G.~J.}\ \bibnamefont {Snyder}},\ }\bibfield  {title}
	{\bibinfo {title} {Solidification processing of alloys in the pseudo-binary
			PbTe - Sb$_2$Te$_3$ system},\ }\href@noop {} {\bibfield  {journal} {\bibinfo
			{journal} {Acta Materialia}\ }\textbf {\bibinfo {volume} {55}},\ \bibinfo
		{pages} {1227} (\bibinfo {year} {2007}{\natexlab{a}})}\BibitemShut {NoStop}%
	\bibitem [{\citenamefont {Shelimova}\ \emph {et~al.}(2007)\citenamefont
		{Shelimova}, \citenamefont {Svechnikova}, \citenamefont {Konstantinov},
		\citenamefont {Karpinskii}, \citenamefont {Avilov}, \citenamefont {Kretova},\
		and\ \citenamefont {Zemskov}}]{Shelimova2007}%
	\BibitemOpen
	\bibfield  {author} {\bibinfo {author} {\bibfnamefont {L.~E.}\ \bibnamefont
			{Shelimova}}, \bibinfo {author} {\bibfnamefont {T.~E.}\ \bibnamefont
			{Svechnikova}}, \bibinfo {author} {\bibfnamefont {P.~P.}\ \bibnamefont
			{Konstantinov}}, \bibinfo {author} {\bibfnamefont {O.~G.}\ \bibnamefont
			{Karpinskii}}, \bibinfo {author} {\bibfnamefont {E.~S.}\ \bibnamefont
			{Avilov}}, \bibinfo {author} {\bibfnamefont {M.~A.}\ \bibnamefont
			{Kretova}},\ and\ \bibinfo {author} {\bibfnamefont {V.~S.}\ \bibnamefont
			{Zemskov}},\ }\bibfield  {title} {\bibinfo {title} {Anisotropic
			thermoelectric properties of the layered compounds PbSb$_2$Te$_4$ and PbBi$_4$Te$_7$},\
	}\href@noop {} {\bibfield  {journal} {\bibinfo  {journal} {Inorganic
				Materials}\ }\textbf {\bibinfo {volume} {43}},\ \bibinfo {pages} {125}
		(\bibinfo {year} {2007})}\BibitemShut {NoStop}%
	\bibitem [{\citenamefont {Jin}\ \emph {et~al.}(2011)\citenamefont {Jin},
		\citenamefont {Song}, \citenamefont {Freeman},\ and\ \citenamefont
		{Kanatzidis}}]{Jin2011}%
	\BibitemOpen
	\bibfield  {author} {\bibinfo {author} {\bibfnamefont {H.}~\bibnamefont
			{Jin}}, \bibinfo {author} {\bibfnamefont {J.-H.}\ \bibnamefont {Song}},
		\bibinfo {author} {\bibfnamefont {A.~J.}\ \bibnamefont {Freeman}},\ and\
		\bibinfo {author} {\bibfnamefont {M.~G.}\ \bibnamefont {Kanatzidis}},\
	}\bibfield  {title} {\bibinfo {title} {Candidates for topological insulators:
			Pb-based chalcogenide series},\ }\href@noop {} {\bibfield  {journal}
		{\bibinfo  {journal} {Phys. Rev. B}\ }\textbf {\bibinfo {volume} {83}},\
		\bibinfo {pages} {041202(R)} (\bibinfo {year} {2011})}\BibitemShut {NoStop}%
	\bibitem [{\citenamefont {Menshchikova}\ \emph {et~al.}(2013)\citenamefont
		{Menshchikova}, \citenamefont {Eremeev},\ and\ \citenamefont
		{Chulkov}}]{MENSHCHIKOVA2013}%
	\BibitemOpen
	\bibfield  {author} {\bibinfo {author} {\bibfnamefont {T.~V.}\ \bibnamefont
			{Menshchikova}}, \bibinfo {author} {\bibfnamefont {S.~V.}\ \bibnamefont
			{Eremeev}},\ and\ \bibinfo {author} {\bibfnamefont {E.~V.}\ \bibnamefont
			{Chulkov}},\ }\bibfield  {title} {\bibinfo {title} {Electronic structure of
			SnSb$_2$Te$_4$ and PbSb$_2$Te$_4$ topological insulators},\ }\href@noop {} {\bibfield
		{journal} {\bibinfo  {journal} {Appl. Surf. Sci.}\ }\textbf {\bibinfo
			{volume} {267}},\ \bibinfo {pages} {1} (\bibinfo {year} {2013})}\BibitemShut
	{NoStop}%
	\bibitem [{\citenamefont {Souma}\ \emph {et~al.}(2012)\citenamefont {Souma},
		\citenamefont {Eto}, \citenamefont {Nomura}, \citenamefont {Nakayama},
		\citenamefont {Sato}, \citenamefont {Takahashi}, \citenamefont {Segawa},\
		and\ \citenamefont {Ando}}]{Souma2012}%
	\BibitemOpen
	\bibfield  {author} {\bibinfo {author} {\bibfnamefont {S.}~\bibnamefont
			{Souma}}, \bibinfo {author} {\bibfnamefont {K.}~\bibnamefont {Eto}}, \bibinfo
		{author} {\bibfnamefont {M.}~\bibnamefont {Nomura}}, \bibinfo {author}
		{\bibfnamefont {K.}~\bibnamefont {Nakayama}}, \bibinfo {author}
		{\bibfnamefont {T.}~\bibnamefont {Sato}}, \bibinfo {author} {\bibfnamefont
			{T.}~\bibnamefont {Takahashi}}, \bibinfo {author} {\bibfnamefont
			{K.}~\bibnamefont {Segawa}},\ and\ \bibinfo {author} {\bibfnamefont
			{Y.}~\bibnamefont {Ando}},\ }\bibfield  {title} {\bibinfo {title}
		{Topological surface states in lead-based ternary telluride
			$\mathrm{Pb}({\mathrm{Bi}}_{1\ensuremath{-}x}{\mathrm{Sb}}_{x}{)}_{2}{\mathrm{Te}}_{4}$},\
	}\href@noop {} {\bibfield  {journal} {\bibinfo  {journal} {Phys. Rev. Lett.}\
		}\textbf {\bibinfo {volume} {108}},\ \bibinfo {pages} {116801} (\bibinfo
		{year} {2012})}\BibitemShut {NoStop}%
	\bibitem [{\citenamefont {Wang}\ \emph {et~al.}(2012)\citenamefont {Wang},
		\citenamefont {Chen}, \citenamefont {Zhu},\ and\ \citenamefont
		{Zhang}}]{Wang2012}%
	\BibitemOpen
	\bibfield  {author} {\bibinfo {author} {\bibfnamefont {J.}~\bibnamefont
			{Wang}}, \bibinfo {author} {\bibfnamefont {X.}~\bibnamefont {Chen}}, \bibinfo
		{author} {\bibfnamefont {B.-F.}\ \bibnamefont {Zhu}},\ and\ \bibinfo {author}
		{\bibfnamefont {S.-C.}\ \bibnamefont {Zhang}},\ }\bibfield  {title} {\bibinfo
		{title} {Topological $p$-$n$ junction},\ }\href@noop {} {\bibfield  {journal}
		{\bibinfo  {journal} {Phys. Rev. B}\ }\textbf {\bibinfo {volume} {85}},\
		\bibinfo {pages} {235131} (\bibinfo {year} {2012})}\BibitemShut {NoStop}%
	\bibitem [{\citenamefont {Hattori}\ \emph {et~al.}(2017)\citenamefont
		{Hattori}, \citenamefont {Tokumoto},\ and\ \citenamefont
		{Edagawa}}]{Hattori2017}%
	\BibitemOpen
	\bibfield  {author} {\bibinfo {author} {\bibfnamefont {Y.}~\bibnamefont
			{Hattori}}, \bibinfo {author} {\bibfnamefont {Y.}~\bibnamefont {Tokumoto}},\
		and\ \bibinfo {author} {\bibfnamefont {K.}~\bibnamefont {Edagawa}},\
	}\bibfield  {title} {\bibinfo {title} {Optimizing composition of
			$\mathrm{Pb}{(\mathrm{B}{\mathrm{i}}_{1\ensuremath{-}x}\mathrm{S}{\mathrm{b}}_{x})}_{2}\mathrm{T}{\mathrm{e}}_{4}$
			topological insulator to achieve a bulk-insulating state},\ }\href@noop {}
	{\bibfield  {journal} {\bibinfo  {journal} {Phys. Rev. Materials}\ }\textbf
		{\bibinfo {volume} {1}},\ \bibinfo {pages} {074201} (\bibinfo {year}
		{2017})}\BibitemShut {NoStop}%
	\bibitem [{\citenamefont {Ikeda}\ \emph
		{et~al.}(2007{\natexlab{b}})\citenamefont {Ikeda}, \citenamefont {Collins},
		\citenamefont {Ravi}, \citenamefont {Gascoin}, \citenamefont {Haile},\ and\
		\citenamefont {Snyder}}]{Ikeda2007B}%
	\BibitemOpen
	\bibfield  {author} {\bibinfo {author} {\bibfnamefont {T.}~\bibnamefont
			{Ikeda}}, \bibinfo {author} {\bibfnamefont {L.~A.}\ \bibnamefont {Collins}},
		\bibinfo {author} {\bibfnamefont {V.~A.}\ \bibnamefont {Ravi}}, \bibinfo
		{author} {\bibfnamefont {F.~S.}\ \bibnamefont {Gascoin}}, \bibinfo {author}
		{\bibfnamefont {S.~M.}\ \bibnamefont {Haile}},\ and\ \bibinfo {author}
		{\bibfnamefont {G.~J.}\ \bibnamefont {Snyder}},\ }\bibfield  {title}
	{\bibinfo {title} {Self-assembled nanometer lamellae of thermoelectric PbTe
			and Sb$_2$Te$_3$ with epitaxy-like interfaces},\ }\href@noop {} {\bibfield
		{journal} {\bibinfo  {journal} {Chemistry of Materials}\ }\textbf {\bibinfo
			{volume} {19}},\ \bibinfo {pages} {763} (\bibinfo {year}
		{2007}{\natexlab{b}})}\BibitemShut {NoStop}%
	\bibitem [{\citenamefont {Fu}(2009)}]{Fu2009}%
	\BibitemOpen
	\bibfield  {author} {\bibinfo {author} {\bibfnamefont {L.}~\bibnamefont
			{Fu}},\ }\bibfield  {title} {\bibinfo {title} {Hexagonal warping effects in
			the surface states of the topological insulator
			${\mathrm{Bi}}_{2}{\mathrm{Te}}_{3}$},\ }\href@noop {} {\bibfield  {journal}
		{\bibinfo  {journal} {Phys. Rev. Lett.}\ }\textbf {\bibinfo {volume} {103}},\
		\bibinfo {pages} {266801} (\bibinfo {year} {2009})}\BibitemShut {NoStop}%
	\bibitem [{Sup()}]{Suppl}%
	\BibitemOpen
	\href@noop {} {}\bibinfo {note} {See Supplemental Material at [URL] for the details of crystal growth and analyses; surface characterization by STM; DFT calculations; QPI simulations; variation of the DP across the surface; symmetrization of QPI patterns; QPI patterns at energies away from the DP; surface-bulk hybridization; QPI simulation including the bulk bands; energy dispersion of the surface states. The Supplemental Material also
contains Refs. \cite{Ozaki2003, Perdew1996}}\BibitemShut {Stop}%
\bibitem [{\citenamefont {Ozaki}(2003)}]{Ozaki2003}%
  \BibitemOpen
  \bibfield  {author} {\bibinfo {author} {\bibfnamefont {T.}~\bibnamefont
  {Ozaki}},\ }\bibfield  {title} {\bibinfo {title} {{Variationally optimized
  atomic orbitals for large-scale electronic structures}},\ }\href@noop {}
  {\bibfield  {journal} {\bibinfo  {journal} {Phys. Rev. B}\ }\textbf
  {\bibinfo {volume} {67}},\ \bibinfo {pages} {155108} (\bibinfo {year}
  {2003})}\BibitemShut {NoStop}%
\bibitem [{\citenamefont {Perdew}\ \emph {et~al.}(1996)\citenamefont {Perdew},
  \citenamefont {Burke},\ and\ \citenamefont {Ernzerhof}}]{Perdew1996}%
  \BibitemOpen
  \bibfield  {author} {\bibinfo {author} {\bibfnamefont {J.~P.}\ \bibnamefont
  {Perdew}}, \bibinfo {author} {\bibfnamefont {K.}~\bibnamefont {Burke}},\ and\
  \bibinfo {author} {\bibfnamefont {M.}~\bibnamefont {Ernzerhof}},\ }\bibfield
  {title} {\bibinfo {title} {{Generalized Gradient Approximation Made
  Simple}},\ }\href@noop {} {\bibfield  {journal} {\bibinfo  {journal}
  {Phys. Rev. Lett.}\ }\textbf {\bibinfo {volume} {77}},\ \bibinfo
  {pages} {3865} (\bibinfo {year} {1996})}\BibitemShut {NoStop}%
	\bibitem{Note}
	\href@noop {} {}\bibinfo {note2}{According to a previously reported XRD analysis of PbSb$_{2}$Te$_{4} $ \cite{Shelimova2004}, Te and Sb sites are slightly substituted with Pb atoms and Pb sites are partially replaced with Sb and Te atoms. In addition, 20\% of the Sb sites were randomly replaced with Bi atoms in our sample. }\BibitemShut {Stop}%
	\bibitem [{\citenamefont {Shelimova}\ \emph {et~al.}(2004)\citenamefont
		{Shelimova}, \citenamefont {Karpinskii}, \citenamefont {Svechnikova},
		\citenamefont {Avilov}, \citenamefont {Kretova},\ and\ \citenamefont
		{Zemskov}}]{Shelimova2004}%
	\BibitemOpen
	\bibfield  {author} {\bibinfo {author} {\bibfnamefont {L.~E.}\ \bibnamefont
			{Shelimova}}, \bibinfo {author} {\bibfnamefont {O.~G.}\ \bibnamefont
			{Karpinskii}}, \bibinfo {author} {\bibfnamefont {T.~E.}\ \bibnamefont
			{Svechnikova}}, \bibinfo {author} {\bibfnamefont {E.~S.}\ \bibnamefont
			{Avilov}}, \bibinfo {author} {\bibfnamefont {M.~A.}\ \bibnamefont
			{Kretova}},\ and\ \bibinfo {author} {\bibfnamefont {V.~S.}\ \bibnamefont
			{Zemskov}},\ }\bibfield  {title} {\bibinfo {title} {Synthesis and structure
			of layered compounds in the PbTe-Bi$_2$Te$_3$ and PbTe-Sb$_2$Te$_3$ systems},\
	}\href@noop {} {\bibfield  {journal} {\bibinfo  {journal} {Inorganic
				Materials}\ }\textbf {\bibinfo {volume} {40}},\ \bibinfo {pages} {1264}
		(\bibinfo {year} {2004})}\BibitemShut {NoStop}%
	\bibitem{Feenstra2005} R. M. Feenstra, S. Gaan, G. Meyer, and K. H. Rieder, Low-temperature tunneling spectroscopy of Ge(111)-c(2 $\times$ 8) surfaces, Phys. Rev. B \textbf{71}, 125316 (2005).
	\bibitem [{\citenamefont {Kuroda}\ \emph {et~al.}(2012)\citenamefont {Kuroda},
		\citenamefont {Miyahara}, \citenamefont {Ye}, \citenamefont {Eremeev},
		\citenamefont {Koroteev}, \citenamefont {Krasovskii}, \citenamefont
		{Chulkov}, \citenamefont {Hiramoto}, \citenamefont {Moriyoshi}, \citenamefont
		{Kuroiwa}, \citenamefont {Miyamoto}, \citenamefont {Okuda}, \citenamefont
		{Arita}, \citenamefont {Shimada}, \citenamefont {Namatame}, \citenamefont
		{Taniguchi}, \citenamefont {Ueda},\ and\ \citenamefont
		{Kimura}}]{Kuroda2012}%
	\BibitemOpen
	\bibfield  {author} {\bibinfo {author} {\bibfnamefont {K.}~\bibnamefont
			{Kuroda}}, \bibinfo {author} {\bibfnamefont {H.}~\bibnamefont {Miyahara}},
		\bibinfo {author} {\bibfnamefont {M.}~\bibnamefont {Ye}}, \bibinfo {author}
		{\bibfnamefont {S.~V.}\ \bibnamefont {Eremeev}}, \bibinfo {author}
		{\bibfnamefont {Y.~M.}\ \bibnamefont {Koroteev}}, \bibinfo {author}
		{\bibfnamefont {E.~E.}\ \bibnamefont {Krasovskii}}, \bibinfo {author}
		{\bibfnamefont {E.~V.}\ \bibnamefont {Chulkov}}, \bibinfo {author}
		{\bibfnamefont {S.}~\bibnamefont {Hiramoto}}, \bibinfo {author}
		{\bibfnamefont {C.}~\bibnamefont {Moriyoshi}}, \bibinfo {author}
		{\bibfnamefont {Y.}~\bibnamefont {Kuroiwa}}, \bibinfo {author} {\bibfnamefont
			{K.}~\bibnamefont {Miyamoto}}, \bibinfo {author} {\bibfnamefont
			{T.}~\bibnamefont {Okuda}}, \bibinfo {author} {\bibfnamefont
			{M.}~\bibnamefont {Arita}}, \bibinfo {author} {\bibfnamefont
			{K.}~\bibnamefont {Shimada}}, \bibinfo {author} {\bibfnamefont
			{H.}~\bibnamefont {Namatame}}, \bibinfo {author} {\bibfnamefont
			{M.}~\bibnamefont {Taniguchi}}, \bibinfo {author} {\bibfnamefont
			{Y.}~\bibnamefont {Ueda}},\ and\ \bibinfo {author} {\bibfnamefont
			{A.}~\bibnamefont {Kimura}},\ }\bibfield  {title} {\bibinfo {title}
		{Experimental verification of ${\mathrm{PbBi}}_{2}{\mathrm{Te}}_{4}$ as a 3D
			topological insulator},\ }\href@noop {} {\bibfield  {journal} {\bibinfo
			{journal} {Phys. Rev. Lett.}\ }\textbf {\bibinfo {volume} {108}},\ \bibinfo
		{pages} {206803} (\bibinfo {year} {2012})}\BibitemShut {NoStop}%
	\bibitem [{\citenamefont {Zhang}\ \emph {et~al.}(2009)\citenamefont {Zhang},
		\citenamefont {Cheng}, \citenamefont {Chen}, \citenamefont {Jia},
		\citenamefont {Ma}, \citenamefont {He}, \citenamefont {Wang}, \citenamefont
		{Zhang}, \citenamefont {Dai}, \citenamefont {Fang}, \citenamefont {Xie},\
		and\ \citenamefont {Xue}}]{Zhang2009}%
	\BibitemOpen
	\bibfield  {author} {\bibinfo {author} {\bibfnamefont {T.}~\bibnamefont
			{Zhang}}, \bibinfo {author} {\bibfnamefont {P.}~\bibnamefont {Cheng}},
		\bibinfo {author} {\bibfnamefont {X.}~\bibnamefont {Chen}}, \bibinfo {author}
		{\bibfnamefont {J.-F.}\ \bibnamefont {Jia}}, \bibinfo {author} {\bibfnamefont
			{X.}~\bibnamefont {Ma}}, \bibinfo {author} {\bibfnamefont {K.}~\bibnamefont
			{He}}, \bibinfo {author} {\bibfnamefont {L.}~\bibnamefont {Wang}}, \bibinfo
		{author} {\bibfnamefont {H.}~\bibnamefont {Zhang}}, \bibinfo {author}
		{\bibfnamefont {X.}~\bibnamefont {Dai}}, \bibinfo {author} {\bibfnamefont
			{Z.}~\bibnamefont {Fang}}, \bibinfo {author} {\bibfnamefont {X.}~\bibnamefont
			{Xie}},\ and\ \bibinfo {author} {\bibfnamefont {Q.-K.}\ \bibnamefont {Xue}},\
	}\bibfield  {title} {\bibinfo {title} {Experimental demonstration of
			topological surface states protected by time-reversal symmetry},\ }\href@noop
	{} {\bibfield  {journal} {\bibinfo  {journal} {Phys. Rev. Lett.}\ }\textbf
		{\bibinfo {volume} {103}},\ \bibinfo {pages} {266803} (\bibinfo {year}
		{2009})}\BibitemShut {NoStop}%
	\bibitem [{\citenamefont {Okada}\ \emph {et~al.}(2011)\citenamefont {Okada},
		\citenamefont {Dhital}, \citenamefont {Zhou}, \citenamefont {Huemiller},
		\citenamefont {Lin}, \citenamefont {Basak}, \citenamefont {Bansil},
		\citenamefont {Huang}, \citenamefont {Ding}, \citenamefont {Wang},
		\citenamefont {Wilson},\ and\ \citenamefont {Madhavan}}]{Okada2011}%
	\BibitemOpen
	\bibfield  {author} {\bibinfo {author} {\bibfnamefont {Y.}~\bibnamefont
			{Okada}}, \bibinfo {author} {\bibfnamefont {C.}~\bibnamefont {Dhital}},
		\bibinfo {author} {\bibfnamefont {W.}~\bibnamefont {Zhou}}, \bibinfo {author}
		{\bibfnamefont {E.~D.}\ \bibnamefont {Huemiller}}, \bibinfo {author}
		{\bibfnamefont {H.}~\bibnamefont {Lin}}, \bibinfo {author} {\bibfnamefont
			{S.}~\bibnamefont {Basak}}, \bibinfo {author} {\bibfnamefont
			{A.}~\bibnamefont {Bansil}}, \bibinfo {author} {\bibfnamefont {Y.-B.}\
			\bibnamefont {Huang}}, \bibinfo {author} {\bibfnamefont {H.}~\bibnamefont
			{Ding}}, \bibinfo {author} {\bibfnamefont {Z.}~\bibnamefont {Wang}}, \bibinfo
		{author} {\bibfnamefont {S.~D.}\ \bibnamefont {Wilson}},\ and\ \bibinfo
		{author} {\bibfnamefont {V.}~\bibnamefont {Madhavan}},\ }\bibfield  {title}
	{\bibinfo {title} {Direct observation of broken time-reversal symmetry on the
			surface of a magnetically doped topological insulator},\ }\href@noop {}
	{\bibfield  {journal} {\bibinfo  {journal} {Phys. Rev. Lett.}\ }\textbf
		{\bibinfo {volume} {106}},\ \bibinfo {pages} {206805} (\bibinfo {year}
		{2011})}\BibitemShut {NoStop}%
	\bibitem [{\citenamefont {Stolyarov}\ \emph {et~al.}(2021)\citenamefont
		{Stolyarov}, \citenamefont {Sheina}, \citenamefont {Khokhlov}, \citenamefont
		{Vlaic}, \citenamefont {Pons}, \citenamefont {Aubin}, \citenamefont
		{Akzyanov}, \citenamefont {Vasenko}, \citenamefont {Menshchikova},
		\citenamefont {Chulkov}, \citenamefont {Golubov}, \citenamefont {Cren},\ and\
		\citenamefont {Roditchev}}]{Stolyarov2021}%
	\BibitemOpen
	\bibfield  {author} {\bibinfo {author} {\bibfnamefont {V.~S.}\ \bibnamefont
			{Stolyarov}}, \bibinfo {author} {\bibfnamefont {V.~A.}\ \bibnamefont
			{Sheina}}, \bibinfo {author} {\bibfnamefont {D.~A.}\ \bibnamefont
			{Khokhlov}}, \bibinfo {author} {\bibfnamefont {S.}~\bibnamefont {Vlaic}},
		\bibinfo {author} {\bibfnamefont {S.}~\bibnamefont {Pons}}, \bibinfo {author}
		{\bibfnamefont {H.}~\bibnamefont {Aubin}}, \bibinfo {author} {\bibfnamefont
			{R.~S.}\ \bibnamefont {Akzyanov}}, \bibinfo {author} {\bibfnamefont {A.~S.}\
			\bibnamefont {Vasenko}}, \bibinfo {author} {\bibfnamefont {T.~V.}\
			\bibnamefont {Menshchikova}}, \bibinfo {author} {\bibfnamefont {E.~V.}\
			\bibnamefont {Chulkov}}, \bibinfo {author} {\bibfnamefont {A.~A.}\
			\bibnamefont {Golubov}}, \bibinfo {author} {\bibfnamefont {T.}~\bibnamefont
			{Cren}},\ and\ \bibinfo {author} {\bibfnamefont {D.}~\bibnamefont
			{Roditchev}},\ }\bibfield  {title} {\bibinfo {title} {Disorder-promoted
			splitting in quasiparticle interference at nesting vectors},\ } {\bibfield  {journal} {\bibinfo
			{journal} {J. Phys. Chem. Lett.}\ }\textbf {\bibinfo
			{volume} {12}},\ \bibinfo {pages} {3127} (\bibinfo {year}
		{2021})}\BibitemShut {NoStop}%
	\bibitem [{\citenamefont {Nurmamat}\ \emph {et~al.}(2013)\citenamefont
		{Nurmamat}, \citenamefont {Krasovskii}, \citenamefont {Kuroda}, \citenamefont
		{Ye}, \citenamefont {Miyamoto}, \citenamefont {Nakatake}, \citenamefont
		{Okuda}, \citenamefont {Namatame}, \citenamefont {Taniguchi}, \citenamefont
		{Chulkov}, \citenamefont {Kokh}, \citenamefont {Tereshchenko},\ and\
		\citenamefont {Kimura}}]{Nurmamat2013}%
	\BibitemOpen
	\bibfield  {author} {\bibinfo {author} {\bibfnamefont {M.}~\bibnamefont
			{Nurmamat}}, \bibinfo {author} {\bibfnamefont {E.~E.}\ \bibnamefont
			{Krasovskii}}, \bibinfo {author} {\bibfnamefont {K.}~\bibnamefont {Kuroda}},
		\bibinfo {author} {\bibfnamefont {M.}~\bibnamefont {Ye}}, \bibinfo {author}
		{\bibfnamefont {K.}~\bibnamefont {Miyamoto}}, \bibinfo {author}
		{\bibfnamefont {M.}~\bibnamefont {Nakatake}}, \bibinfo {author}
		{\bibfnamefont {T.}~\bibnamefont {Okuda}}, \bibinfo {author} {\bibfnamefont
			{H.}~\bibnamefont {Namatame}}, \bibinfo {author} {\bibfnamefont
			{M.}~\bibnamefont {Taniguchi}}, \bibinfo {author} {\bibfnamefont {E.~V.}\
			\bibnamefont {Chulkov}}, \bibinfo {author} {\bibfnamefont {K.~A.}\
			\bibnamefont {Kokh}}, \bibinfo {author} {\bibfnamefont {O.~E.}\ \bibnamefont
			{Tereshchenko}},\ and\ \bibinfo {author} {\bibfnamefont {A.}~\bibnamefont
			{Kimura}},\ }\bibfield  {title} {\bibinfo {title} {Unoccupied topological
			surface state in Bi${}_{2}$Te${}_{2}$Se},\ }\href@noop {} {\bibfield
		{journal} {\bibinfo  {journal} {Phys. Rev. B}\ }\textbf {\bibinfo {volume}
			{88}},\ \bibinfo {pages} {081301(R)} (\bibinfo {year} {2013})}\BibitemShut
	{NoStop}%
	\bibitem [{\citenamefont {Kohsaka}\ \emph {et~al.}(2017)\citenamefont
		{Kohsaka}, \citenamefont {Machida}, \citenamefont {Iwaya}, \citenamefont
		{Kanou}, \citenamefont {Hanaguri},\ and\ \citenamefont
		{Sasagawa}}]{Kohsaka2017}%
	\BibitemOpen
	\bibfield  {author} {\bibinfo {author} {\bibfnamefont {Y.}~\bibnamefont
			{Kohsaka}}, \bibinfo {author} {\bibfnamefont {T.}~\bibnamefont {Machida}},
		\bibinfo {author} {\bibfnamefont {K.}~\bibnamefont {Iwaya}}, \bibinfo
		{author} {\bibfnamefont {M.}~\bibnamefont {Kanou}}, \bibinfo {author}
		{\bibfnamefont {T.}~\bibnamefont {Hanaguri}},\ and\ \bibinfo {author}
		{\bibfnamefont {T.}~\bibnamefont {Sasagawa}},\ }\bibfield  {title} {\bibinfo
		{title} {Spin-orbit scattering visualized in quasiparticle interference},\
	}\href@noop {} {\bibfield  {journal} {\bibinfo  {journal} {Phys. Rev. B}\
		}\textbf {\bibinfo {volume} {95}},\ \bibinfo {pages} {115307} (\bibinfo
		{year} {2017})}\BibitemShut {NoStop}%
	\bibitem [{\citenamefont {Wang}\ \emph {et~al.}(2011)\citenamefont {Wang},
		\citenamefont {Li}, \citenamefont {Cheng}, \citenamefont {Song},
		\citenamefont {Zhang}, \citenamefont {Deng}, \citenamefont {Chen},
		\citenamefont {Ma}, \citenamefont {He}, \citenamefont {Jia}, \citenamefont
		{Xue},\ and\ \citenamefont {Zhu}}]{Wang2011}%
	\BibitemOpen
	\bibfield  {author} {\bibinfo {author} {\bibfnamefont {J.}~\bibnamefont
			{Wang}}, \bibinfo {author} {\bibfnamefont {W.}~\bibnamefont {Li}}, \bibinfo
		{author} {\bibfnamefont {P.}~\bibnamefont {Cheng}}, \bibinfo {author}
		{\bibfnamefont {C.}~\bibnamefont {Song}}, \bibinfo {author} {\bibfnamefont
			{T.}~\bibnamefont {Zhang}}, \bibinfo {author} {\bibfnamefont
			{P.}~\bibnamefont {Deng}}, \bibinfo {author} {\bibfnamefont {X.}~\bibnamefont
			{Chen}}, \bibinfo {author} {\bibfnamefont {X.}~\bibnamefont {Ma}}, \bibinfo
		{author} {\bibfnamefont {K.}~\bibnamefont {He}}, \bibinfo {author}
		{\bibfnamefont {J.-F.}\ \bibnamefont {Jia}}, \bibinfo {author} {\bibfnamefont
			{Q.-K.}\ \bibnamefont {Xue}},\ and\ \bibinfo {author} {\bibfnamefont {B.-F.}\
			\bibnamefont {Zhu}},\ }\bibfield  {title} {\bibinfo {title} {Power-law decay
			of standing waves on the surface of topological insulators},\ }\href@noop {}
	{\bibfield  {journal} {\bibinfo  {journal} {Phys. Rev. B}\ }\textbf {\bibinfo
			{volume} {84}},\ \bibinfo {pages} {235447} (\bibinfo {year}
		{2011})}\BibitemShut {NoStop}%
	\bibitem [{Not()}]{Note1}%
	\BibitemOpen
	\href@noop {} {}\bibinfo {note} {A possible reason for this discrepancy in
		the scale is that, in the QPI simulation, the electronic structure
		calculated for PbSb$_2$Te$_4$ was used as an alternative to the electronic structure of our actual sample
		of Pb(Bi$_{0.20}$Sb$_{0.80}$)$_2$Te$_4$. The detailed band shape for
		PbSb$_2$Te$_4$ should differ from that of Pb(Bi$_{0.20}$Sb$_{0.80}$)$_2$Te$_4$
		because, with an increase of the composition ratio of Bi, the electronic
		structure of the compound should become similar to that of PbBi$_2$Te$_4$
		whose top of the BVB appears in the $\overline{\Gamma} \, \overline{M}$ line
		\cite{Souma2012,Kuroda2012}.}\BibitemShut {Stop}%
	\bibitem [{\citenamefont {Li}\ and\ \citenamefont {Carbotte}(2014)}]{Li2014}%
	\BibitemOpen
	\bibfield  {author} {\bibinfo {author} {\bibfnamefont {Z.}~\bibnamefont
			{Li}}\ and\ \bibinfo {author} {\bibfnamefont {J.~P.}\ \bibnamefont
			{Carbotte}},\ }\bibfield  {title} {\bibinfo {title} {Hexagonal warping on
			spin texture, Hall conductivity, and circular dichroism of topological
			insulators},\ }\href@noop {} {\bibfield  {journal} {\bibinfo  {journal}
			{Phys. Rev. B}\ }\textbf {\bibinfo {volume} {89}},\ \bibinfo {pages} {165420}
		(\bibinfo {year} {2014})}\BibitemShut {NoStop}%
	\bibitem [{\citenamefont {Akzyanov}\ and\ \citenamefont
		{Rakhmanov}(2018)}]{Akzyanov2018}%
	\BibitemOpen
	\bibfield  {author} {\bibinfo {author} {\bibfnamefont {R.~S.}\ \bibnamefont
			{Akzyanov}}\ and\ \bibinfo {author} {\bibfnamefont {A.~L.}\ \bibnamefont
			{Rakhmanov}},\ }\bibfield  {title} {\bibinfo {title} {Surface charge
			conductivity of a topological insulator in a magnetic field: The effect of
			hexagonal warping},\ }\href@noop {} {\bibfield  {journal} {\bibinfo
			{journal} {Phys. Rev. B}\ }\textbf {\bibinfo {volume} {97}},\ \bibinfo
		{pages} {075421} (\bibinfo {year} {2018})}\BibitemShut {NoStop}%
	\bibitem [{\citenamefont {Okada}\ \emph {et~al.}(2012)\citenamefont {Okada},
		\citenamefont {Zhou}, \citenamefont {Dhital}, \citenamefont {Walkup},
		\citenamefont {Ran}, \citenamefont {Wang}, \citenamefont {Wilson},\ and\
		\citenamefont {Madhavan}}]{Okada2012}%
	\BibitemOpen
	\bibfield  {author} {\bibinfo {author} {\bibfnamefont {Y.}~\bibnamefont
			{Okada}}, \bibinfo {author} {\bibfnamefont {W.}~\bibnamefont {Zhou}},
		\bibinfo {author} {\bibfnamefont {C.}~\bibnamefont {Dhital}}, \bibinfo
		{author} {\bibfnamefont {D.}~\bibnamefont {Walkup}}, \bibinfo {author}
		{\bibfnamefont {Y.}~\bibnamefont {Ran}}, \bibinfo {author} {\bibfnamefont
			{Z.}~\bibnamefont {Wang}}, \bibinfo {author} {\bibfnamefont {S.~D.}\
			\bibnamefont {Wilson}},\ and\ \bibinfo {author} {\bibfnamefont
			{V.}~\bibnamefont {Madhavan}},\ }\bibfield  {title} {\bibinfo {title}
		{Visualizing Landau levels of dirac electrons in a one-dimensional
			potential},\ }\href@noop {} {\bibfield  {journal} {\bibinfo  {journal} {Phys.
				Rev. Lett.}\ }\textbf {\bibinfo {volume} {109}},\ \bibinfo {pages} {166407}
		(\bibinfo {year} {2012})}\BibitemShut {NoStop}%
	\bibitem [{\citenamefont {Steinberg}\ \emph {et~al.}(2011)\citenamefont
		{Steinberg}, \citenamefont {Lalo\"e}, \citenamefont {Fatemi}, \citenamefont
		{Moodera},\ and\ \citenamefont {Jarillo-Herrero}}]{Steinberg2011}%
	\BibitemOpen
	\bibfield  {author} {\bibinfo {author} {\bibfnamefont {H.}~\bibnamefont
			{Steinberg}}, \bibinfo {author} {\bibfnamefont {J.-B.}\ \bibnamefont
			{Lalo\"e}}, \bibinfo {author} {\bibfnamefont {V.}~\bibnamefont {Fatemi}},
		\bibinfo {author} {\bibfnamefont {J.~S.}\ \bibnamefont {Moodera}},\ and\
		\bibinfo {author} {\bibfnamefont {P.}~\bibnamefont {Jarillo-Herrero}},\
	}\bibfield  {title} {\bibinfo {title} {Electrically tunable surface-to-bulk
			coherent coupling in topological insulator thin films},\ }\href@noop {}
	{\bibfield  {journal} {\bibinfo  {journal} {Phys. Rev. B}\ }\textbf {\bibinfo
			{volume} {84}},\ \bibinfo {pages} {233101} (\bibinfo {year}
		{2011})}\BibitemShut {NoStop}%
	\bibitem [{\citenamefont {Tian}\ \emph {et~al.}(2014)\citenamefont {Tian},
		\citenamefont {Chang}, \citenamefont {Cao}, \citenamefont {He}, \citenamefont
		{Ma}, \citenamefont {Xue},\ and\ \citenamefont {Chen}}]{Tian2014}%
	\BibitemOpen
	\bibfield  {author} {\bibinfo {author} {\bibfnamefont {J.}~\bibnamefont
			{Tian}}, \bibinfo {author} {\bibfnamefont {C.}~\bibnamefont {Chang}},
		\bibinfo {author} {\bibfnamefont {H.}~\bibnamefont {Cao}}, \bibinfo {author}
		{\bibfnamefont {K.}~\bibnamefont {He}}, \bibinfo {author} {\bibfnamefont
			{X.}~\bibnamefont {Ma}}, \bibinfo {author} {\bibfnamefont {Q.}~\bibnamefont
			{Xue}},\ and\ \bibinfo {author} {\bibfnamefont {Y.~P.}\ \bibnamefont
			{Chen}},\ }\bibfield  {title} {\bibinfo {title} {Quantum and classical
			magnetoresistance in ambipolar topological insulator transistors with
			gate-tunable bulk and surface conduction},\ }\href@noop {} {\bibfield
		{journal} {\bibinfo  {journal} {Scientific Reports}\ }\textbf {\bibinfo
			{volume} {4}},\ \bibinfo {pages} {4859} (\bibinfo {year} {2014})}\BibitemShut
	{NoStop}%
	\bibitem [{\citenamefont {Kondou}\ \emph {et~al.}(2016)\citenamefont {Kondou},
		\citenamefont {Yoshimi}, \citenamefont {Tsukazaki}, \citenamefont {Fukuma},
		\citenamefont {Matsuno}, \citenamefont {Takahashi}, \citenamefont {Kawasaki},
		\citenamefont {Tokura},\ and\ \citenamefont {Otani}}]{Kondou2016}%
	\BibitemOpen
	\bibfield  {author} {\bibinfo {author} {\bibfnamefont {K.}~\bibnamefont
			{Kondou}}, \bibinfo {author} {\bibfnamefont {R.}~\bibnamefont {Yoshimi}},
		\bibinfo {author} {\bibfnamefont {A.}~\bibnamefont {Tsukazaki}}, \bibinfo
		{author} {\bibfnamefont {Y.}~\bibnamefont {Fukuma}}, \bibinfo {author}
		{\bibfnamefont {J.}~\bibnamefont {Matsuno}}, \bibinfo {author} {\bibfnamefont
			{K.~S.}\ \bibnamefont {Takahashi}}, \bibinfo {author} {\bibfnamefont
			{M.}~\bibnamefont {Kawasaki}}, \bibinfo {author} {\bibfnamefont
			{Y.}~\bibnamefont {Tokura}},\ and\ \bibinfo {author} {\bibfnamefont
			{Y.}~\bibnamefont {Otani}},\ }\bibfield  {title} {\bibinfo {title}
		{Fermi-level-dependent charge-to-spin current conversion by Dirac surface
			states of topological insulators},\ }\href@noop {} {\bibfield  {journal}
		{\bibinfo  {journal} {Nature Physics}\ }\textbf {\bibinfo {volume} {12}},\
		\bibinfo {pages} {1027} (\bibinfo {year} {2016})}\BibitemShut {NoStop}%
	\bibitem [{\citenamefont {Khang}\ \emph {et~al.}(2018)\citenamefont {Khang},
		\citenamefont {Ueda},\ and\ \citenamefont {Hai}}]{Khang2018}%
	\BibitemOpen
	\bibfield  {author} {\bibinfo {author} {\bibfnamefont {N.~H.~D.}\
			\bibnamefont {Khang}}, \bibinfo {author} {\bibfnamefont {Y.}~\bibnamefont
			{Ueda}},\ and\ \bibinfo {author} {\bibfnamefont {P.~N.}\ \bibnamefont
			{Hai}},\ }\bibfield  {title} {\bibinfo {title} {A conductive topological
			insulator with large spin Hall effect for ultralow power spin orbit torque
			switching},\ }\href@noop {} {\bibfield  {journal} {\bibinfo  {journal}
			{Nature Materials}\ }\textbf {\bibinfo {volume} {17}},\ \bibinfo {pages}
		{808} (\bibinfo {year} {2018})}\BibitemShut {NoStop}%
\end{thebibliography}
%

\end{document}


\title{Supplemental Materials for "Topological surface states hybridized with bulk states of Bi-doped PbSb$_2$Te$_4$ revealed in quasiparticle interference"}

\author{Yuya Hattori}
\email{HATTORI.Yuya@nims.go.jp}
\affiliation{Research Center for Materials Nanoarchitectonics (MANA), National Institute for Materials Science, 3-13 Sakura, Tsukuba, Ibaraki 305-0003, Japan}
\affiliation{Institute of Industrial Science, The University of Tokyo, Komaba, Meguro-ku, Tokyo 153-8505, Japan}
\author{Keisuke Sagisaka}
\email{SAGISAKA.Keisuke@nims.go.jp}
\affiliation{Center for Basic Research on Materials, National Institute for Materials Science, 1-2-1 Sengen, Tsukuba, Ibaraki 305-0047, Japan.}
\author{Shunsuke Yoshizawa}
\affiliation{Center for Basic Research on Materials, National Institute for Materials Science, 1-2-1 Sengen, Tsukuba, Ibaraki 305-0047, Japan.}
\author{Yuki Tokumoto}
\affiliation{Institute of Industrial Science, The University of Tokyo, Komaba, Meguro-ku, Tokyo 153-8505, Japan}
\author{Keiichi Edagawa}
\affiliation{Institute of Industrial Science, The University of Tokyo, Komaba, Meguro-ku, Tokyo 153-8505, Japan}

\maketitle
\renewcommand{\thefigure}{S\arabic{figure}}
\renewcommand{\thetable}{S\arabic{table}}
\clearpage

\section{Crystal growth and analyses}
A single crystal of Bi-doped PbSb$_2$Te$_4$ [Pb(Bi$_{1-x}$Sb$_{x}$)$_{2}$Te$_{4}$ ($x$ = 0.80)] was grown by the Bridgman method \cite{Hattori2017}. The crystal of PbSb$_2$Te$_4$ consists of seven-layer (SL) units layered by van der Waals forces; thus, it can be easily cleaved at the (0001) plane with Te termination. 

The composition of the synthesized crystals was quantitatively analyzed using an electron-probe microanalyzer. A melt-spun PbBiSbTe$_4$ sample produced by a single-roller method was used as a standard sample for ZAF conversion. The crystalline phases were identified by powder X-ray diffraction (pXRD) analysis using Cu K$\alpha$ radiation. Figure S1(a) shows the batch position dependence of the Sb ratio $x$ = [Sb]/([Sb]+[Bi]) in the Pb(Bi,Sb)$_2$Te$_4$ single phase region. Along the growth direction, $x$ increases to the end point of crystallization. The averaged composition is displayed in Table \ref{tab:tableS1}. As a previous study of PbSb$_2$Te$_4$ showed \cite{Ikeda2007}, the composition of Pb(Bi$_{0.20}$Sb$_{0.80}$)$_2$Te$_4$ was found to differ from its stoichiometric ratio (Pb : Bi+Sb : Te = 14.3 : 28.6 : 57.1), reflecting its disordered nature \cite{Shelimova2004}. However, its crystal structure remained that of PbBi$_2$Te$_4$. In the present study, the Rietveld analysis results in Fig. S1(b) show a good fit with the PbBi$_2$Te$_4$ crystal structure.

\begin{table}[h]
	\caption{\label{tab:tableS1} Averaged composition in the Pb(Bi$_{1-x}$Sb$_x$)$_2$Te$_4$ single phase region.}
	\begin{ruledtabular}
		\begin{tabular}{lcccc}
			& Pb & Bi & Sb & Te\\
			\hline
			atomic ratio (\%) & 10.1	&5.74	&23.2	&61.0 \\
		\end{tabular}
	\end{ruledtabular}
\end{table} 

\begin{figure}[b]
	\begin{center}
		\includegraphics[width=16cm]{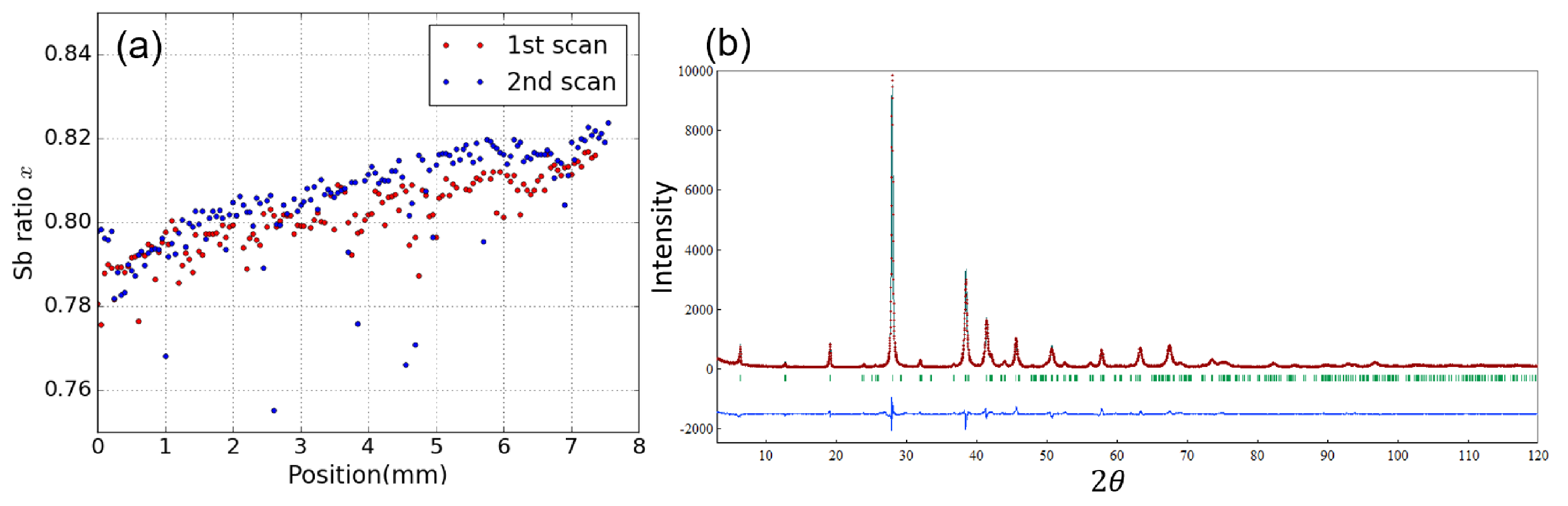}
		\caption{(a) Sb ratio $x$  along the growth direction and (b) the result of pXRD analysis by the Rietveld method for a Pb(Bi$_{1-x}$Sb$_x$)$_2$Te$_4$ (x = 0.8) single phase region.}
		\label{fig:S1}
	\end{center}
\end{figure}

\clearpage

\section{Surface characterization by STM}
The surface characterization was conducted in an ultrahigh-vacuum (UHV) low-temperature STM system (customized Unisoku USM-1300). All STM data were acquired at 4.5 K with a Pt-Ir tip after a clean surface was obtained by cleavage under UHV. The differential conductance (d$I$/d$V$) signals were recorded through a lock-in technique with a modulation voltage of 3 mV at a frequency of 830 Hz.

\section{DFT calculations}
The surface and bulk band structures of PbSb$_{2}$Te$_{4}$ and PbBi$_{2}$Te$_{4}$ were computed using the OpenMX code \cite{Ozaki2003} based on DFT within the Perdew--Burke--Ernzerhof generalized gradient approximation \cite{Perdew1996}. We used pseudo atomic orbitals of Te7.0-$s$3$p$2$d$2$f$1, Sb7.0-$s$3$p$2$d$2$f$1, Bi8.0-$s$3$p$2$d$2$f$1, and Pb8.0-$s$3$p$2$d$2$f$1 as the basis functions, along with fully relativistic pseudopotentials provided in the database of Version 2019. For the computation for bulk bands, we used an energy cutoff of 300 Ry and a 12 $\times$ 12 $\times$ 3 $k$-point mesh. The geometry optimization was conducted until the residual forces were reduced to 1 $\times$ 10$^{5}$ Hartree/Bohr without the spin--orbit coupling. The bulk bands projected to the surface Brillouin zone were then computed with the spin--orbit coupling. For the calculation of the surface states, we used a slab model consisting of three unit cells (nine SLs) and a 4.5 nm thick vacuum layer. We applied an energy cutoff of 300 Ry and a 9 $\times$ 9 $\times$ 1 $k$-point mesh for the slab calculations. These calculations provided the band structure of PbSb$_{2}$Te$_{4}$ projected to the surface [Fig. 1(b) and Fig. S2(a)] and that of PbBi$_{2}$Te$_{4}$ [Fig. S2(b)], which agree well with the result of previously reported band calculations \cite{MENSHCHIKOVA2013, Kuroda2012}.
\begin{figure}[h]
	\begin{center}
		\includegraphics[width=13cm]{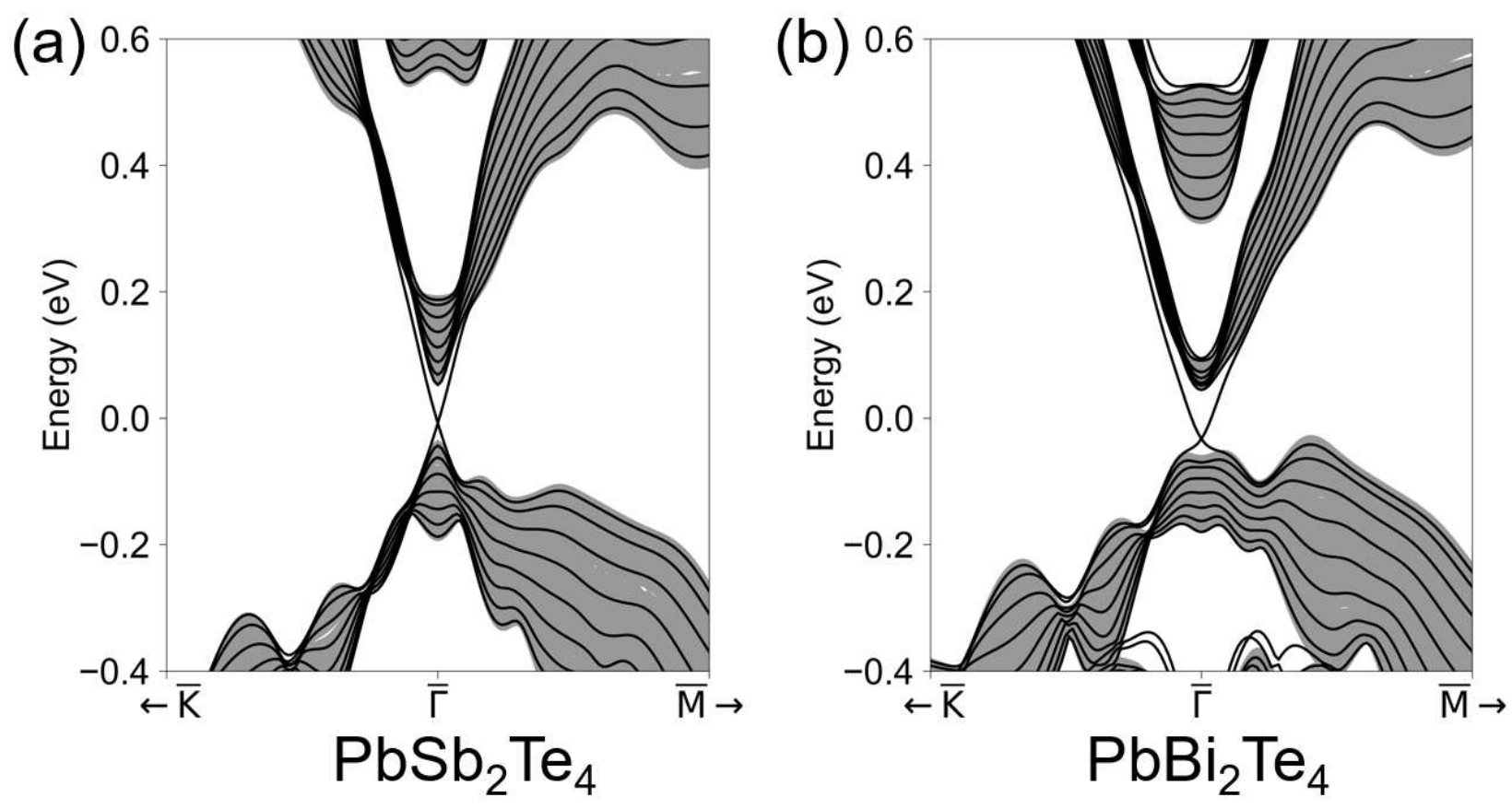}
		\caption{Band diagrams calculated using slab models. (a) PbSb$_2$Te$_4$, (b)PbBi$_2$Te$_4$.}
		\label{fig:S2}
	\end{center}
\end{figure}

\newpage

\section{QPI simulations}
To clarify the formation process of the QPI patterns, we conducted a numerical QPI simulation based on the $T$-matrix formalism \cite{Kohsaka2017}. To the first order of scattering potential $\hat{V}_{\textbf{\textit{k}},\textbf{\textit{k}}'}$, the QPI pattern in the reciprocal space with scattering wave vector $\textbf{\textit{q}} = \textbf{\textit{k}} - \textbf{\textit{k}}'$ is given by
\begin{equation}
	\begin{split}
		\rho(\textbf{\textit{q}},\omega) =  -\frac{1}{\pi}\sum_{\textbf{\textit{k}}}\sum_{m,n}\rm{Im}
			\{\textit{g}_n(\textbf{\textit{k}},\omega) \textit{g}_m(\textbf{\textit{k}}',\omega)\}\\
		\times\braket{\psi_n(\textbf{\textit{k}})|\hat{V}_{\textbf{\textit{k}},\textbf{\textit{k}}'}|\psi_m(\textbf{\textit{k}}')}\braket{\psi_m(\textbf{\textit{k}}')|\psi_n(\textbf{\textit{k}})}
	\end{split}
\end{equation}
where $\textit{g}(\textbf{\textit{k}},\omega)$ = 1/[$\omega$ -- $E_n$($\textbf{\textit{k}}$) + i$\eta$] is Green's function with a broadening factor $\eta$, $\ket{\psi_n(\textbf{\textit{k}})}$ is the $n$-th eigenstate, and $E_n$($\textbf{\textit{k}}$) is the corresponding eigenvalue. We assumed a scalar, momentum-independent scattering potential, $ \hat{V}_{\mathbf{k}} = V_0 I$ ($I$ is the identity matrix) with $V_0$ = 0.1 eV. For the QPI simulation, we produced surface DOS from the results of DFT calculations. We computed the Mulliken population $\rho_n^{\mathrm{surf}}(\textbf{\textit{k}})$ and the spin polarization $\mathbf{s}_n^{\mathrm{surf}}(\textbf{\textit{k}})$ of the pseudo atomic orbitals belonging to the Te atom on the surface. The $|\psi_n(\textbf{\textit{k}})\rangle$ used in the QPI simulation was then constructed as
\begin{equation}
\ket{\psi_n(\textbf{\textit{k}})} 
= \sqrt{(1/2)\rho_n^{\mathrm{surf}}(\textbf{\textit{k}})} 
\left( \cos\frac{\theta(\textbf{\textit{k}})}{2},~\sin\frac{\theta(\textbf{\textit{k}})}{2} e^{i\varphi(\textbf{\textit{k}})} \right)^T \
\end{equation}
where $\theta(\textbf{\textit{k}})$ and $\varphi(\textbf{\textit{k}})$ are the azimuthal and polar angles, respectively, of $\mathbf{s}_n^{\mathrm{surf}}(\textbf{\textit{k}})$. 
The surface DOS in the reciprocal space was computed as $-(2/\pi) \sum_n  \rho_n^{\mathrm{surf}}(\mathbf{k}) \mathrm{Im} \{g_n(\textbf{\textit{k}},\omega)\}$ for different energies.

\section{Variation of the Dirac point across the surface}
\begin{figure}[h]
	\begin{center}
		\includegraphics[width=16cm]{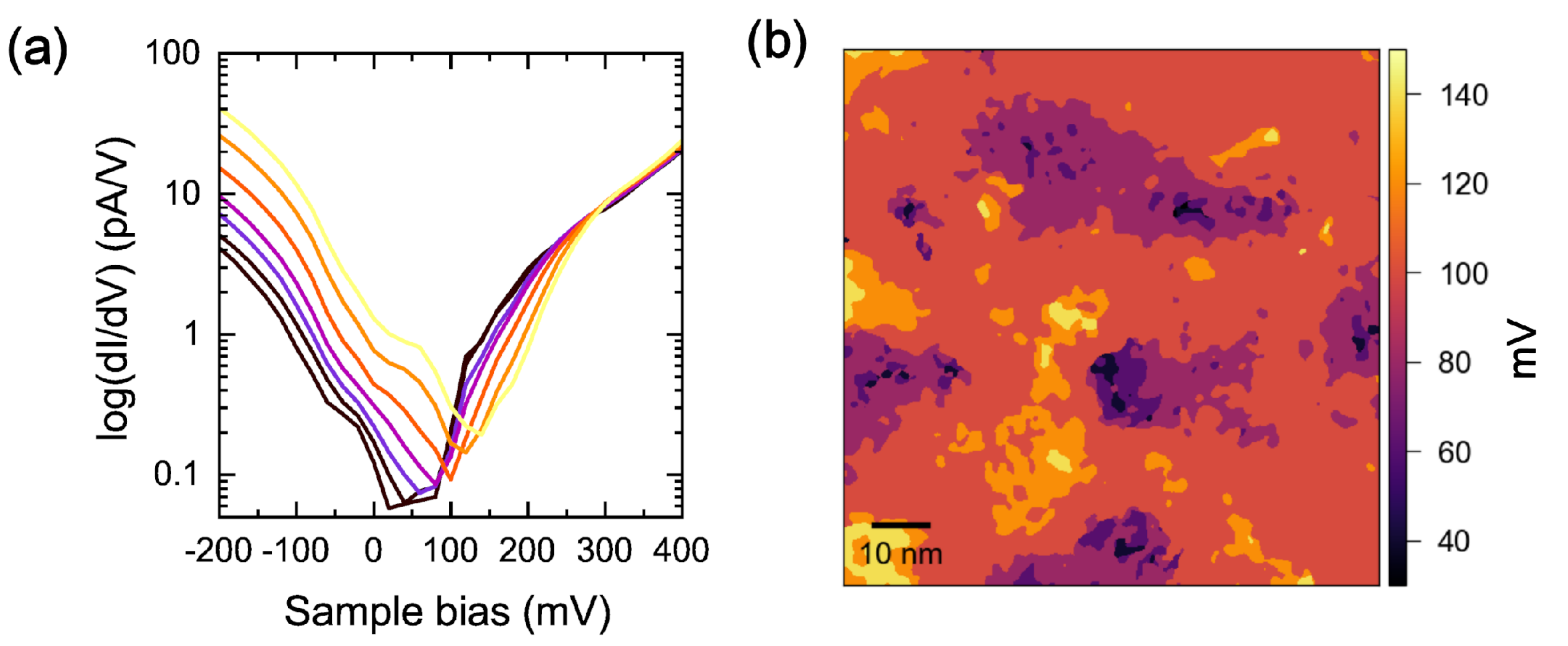} 
		\caption{(a) Surface variation of d$I$/d$V$ spectra. Each spectrum was obtained by averaging the spectra whose differential conductance minima (Dirac point) appeared at the same energy. (b) Distribution of the minima in the d$I$/d$V$ spectra. The color of each spectrum in (a) corresponds to the color map in (b).}
		\label{fig:S3}
	\end{center}
\end{figure}

The energy of the Dirac point (DP) slightly varies from position to position in this surface. Figure \ref{fig:S3} shows the variation of d$I$/d$V$ spectra recorded in an area of 100 $\times$ 100 nm$^2$. The minima of the d$I$/d$V$ curves, corresponding to the Dirac point, range from $V_\textrm{s}$ = 50 to 150 mV; however, the majority of the minima (67\%) appear at $V_\textrm{s}$ = 100 meV. The distribution of the minima (Dirac point) in the surface is displayed in Fig. \ref{fig:S3}. Although the origin of this variation is unknown, it may be related to the variation of the elemental composition in the crystal. 

\newpage

\section{Symmetrization of QPI patterns}
\begin{figure}[h]
	\begin{center}
		\includegraphics[width=16cm]{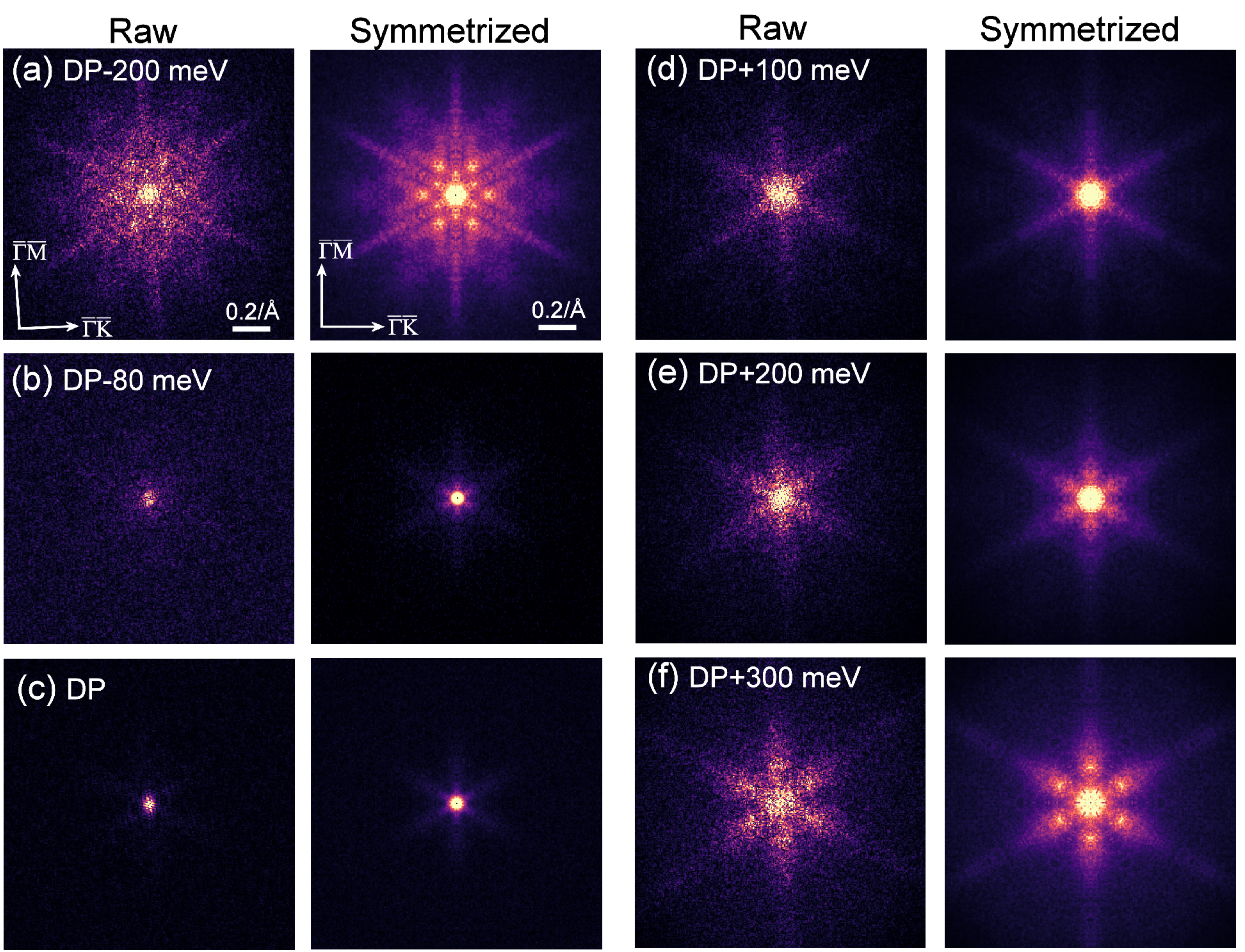} 
		\caption{Comparison of raw and symmetrized Fourier transform image of d$I$/d$V$ recorded at $V_\textrm{s}$ = (a) $-$100, (b) +20, (c) +100, (d) +200, (e) +300, and (f) +400 mV.}
		\label{fig:S4}
	\end{center}
\end{figure}

The QPI patterns presented in Fig. 2 of the main text underwent drift-corrected and sixfold-symmetrized to enhance the signal-to-noise ratio. Figure \ref{fig:S4} shows raw and symmetrized QPI patterns. We have confirmed that the process of symmetrization does not generate any artifacts and effectively enhances the features of QPI patterns that we discuss in the main text. 

\newpage

\section{QPI patterns at energies away from the Dirac point}
\begin{figure}[h]
	\begin{center}
		\includegraphics[width=16cm]{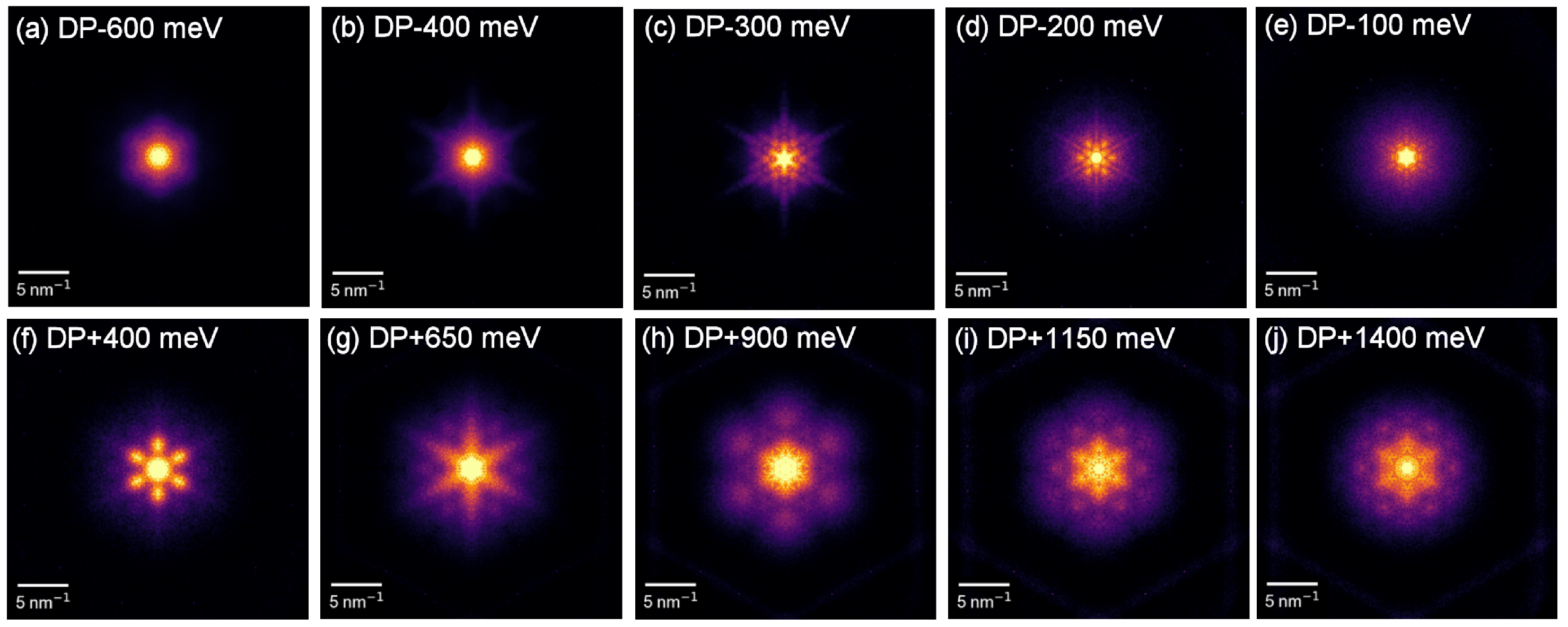} 
		\caption{Fourier transform of d$I$/d$V$ images recorded at $V_\textrm{s}$ = (a) $-$500, (b) $-$300, (c) $-$200, (d) $-$100, (e) 0, (f)+500, (g) +750, (h) +1000, (i) +1250, and (j) +1500 mV.}
		\label{fig:S5}
	\end{center}
\end{figure}

QPI patterns can be imaged at energies away from the DP ($V_\textrm{s}$ = +100 mV). Figure \ref{fig:S5} shows the Fourier transform of d$I$/d$V$ images of the Bi-doped PbSb$_2$Te$_4$ surface at various high sample biases. These QPI images in the reciprocal space differ substantially from those expected for the pure Dirac cone [Fig. 4(a)]. For the energies below the DP, the QPI patterns due to states hybridized between the surface and bulk valence bands were observed at energies as much as 300 meV below the DP [Figs. \ref{fig:S5}(c) and \ref{fig:S5}(d)]. At even lower energies, a distinct QPI pattern was no longer observed, where the surface states were considered to have almost disappeared [Figs. \ref{fig:S5}(a) and \ref{fig:S5}(b)]. In the case of the energies above the DP, the peak feature appearing along the $\overline{\Gamma} \, \overline{M}$ directions derived from the segmentation of the Dirac cone [Fig. 3(d) in the main text] was observed at energies as much as 400 meV above the DP [Fig. \ref{fig:S5}(f)]. At even higher energies, the Fourier transformed QPI patterns exhibited different features than those originating from the Dirac cone. The QPI pattern at $\sim$650 meV above the DP was similar to the patterns observed at 200 meV below the DP. Such a pattern should predominantly reflect bulk conduction bands developing in the $\overline{\Gamma} \, \overline{M}$ line approximately 500 meV above the DP [Fig. \ref{fig:S2}]. At even higher energies, other surface states or surface resonance were imaged as QPI patterns by STM [Figs. \ref{fig:S5}(h)--\ref{fig:S5}(j)] .

\newpage

\section{Surface-bulk hybridization}
\begin{figure}[h]
	\begin{center}
		\includegraphics[width=11cm]{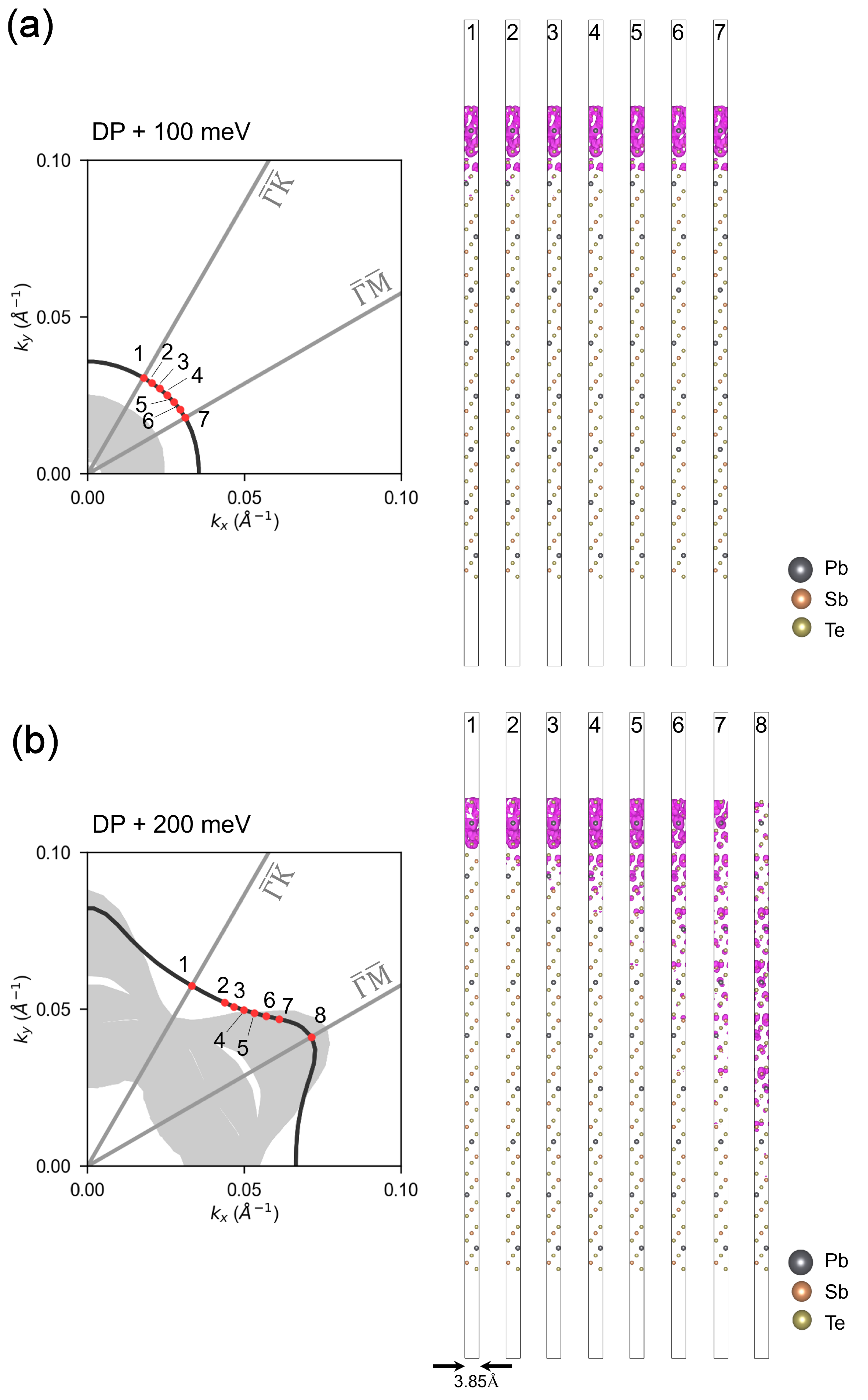} 
		\caption{Spatial distribution of charge densities of the surface band (Dirac cone) in a  PbSb$_2$Te$_4$ slab. (a) 100 meV and (b) 200 meV above the DP. (Left panel) isoenergy contour of the surface band and bulk bands projected to the surface. Gray regions correspond to the bulk bands. (Right panel) Charge density plots of the Dirac cone at selected $k$-points denoted by numerical labels on the black curve in the isoenergy contour. The isosurfaces, depicted in pink, represent a value of 2.5 $\times$ 10$^{-4}$ Bohr$^{-3}$.}
		\label{fig:S6}
	\end{center}
\end{figure}

Figure 3 in the main text shows that the Dirac cone partially disappears around the $\overline{\Gamma} \, \overline{M}$ line at the energies where the bulk conduction bands overlap. To observe this effect visually, we present the spatial distributions of charge densities for the surface band at various $k$-points (the square of the absolute value of the wave functions of the selected band at the selected $k$-point), as illustrated in Fig.  \ref{fig:S6}. At an energy of +100 meV above the Dirac point, the surface state exhibits circular symmetry and sufficiently separated from the bulk bands. Consequently, charges remain localized in the surface at any $k$-points along the surface state. On the other hand, at +200 meV above the Dirac point, the surface state is warped and partially overlaps with the bulk conduction bands around the $\overline{\Gamma} \, \overline{M}$ line. Charges are localized in the surface around the $\overline{\Gamma} \, \overline{K}$ line ($k$-points 1 -- 5). Meanwhile, charges are distributed in the middle of the slab near the $\overline{\Gamma} \, \overline{M}$ line ($k$-points 6 -- 8), where the bulk conduction bands overlap. Around the $\overline{\Gamma} \, \overline{M}$ line, a hybridization of surface and bulk wave functions leads to redistribute charges between the surface and bulk bands. This phenomenon results in a reduction in the Mulliken population of the topmost Te atoms around the $\overline{\Gamma} \, \overline{M}$ line. As the Mulliken population serves as a weighting factor in the computation of the surface DOS, the resultant surface DOS experiences truncation at the vertices of the Dirac cone, as displayed in Figs. 3(c) and 3(d) in the main text.

\clearpage

\section{QPI simulation including the bulk bands}
\begin{figure}[h]
	\begin{center}
		\includegraphics[width=16cm]{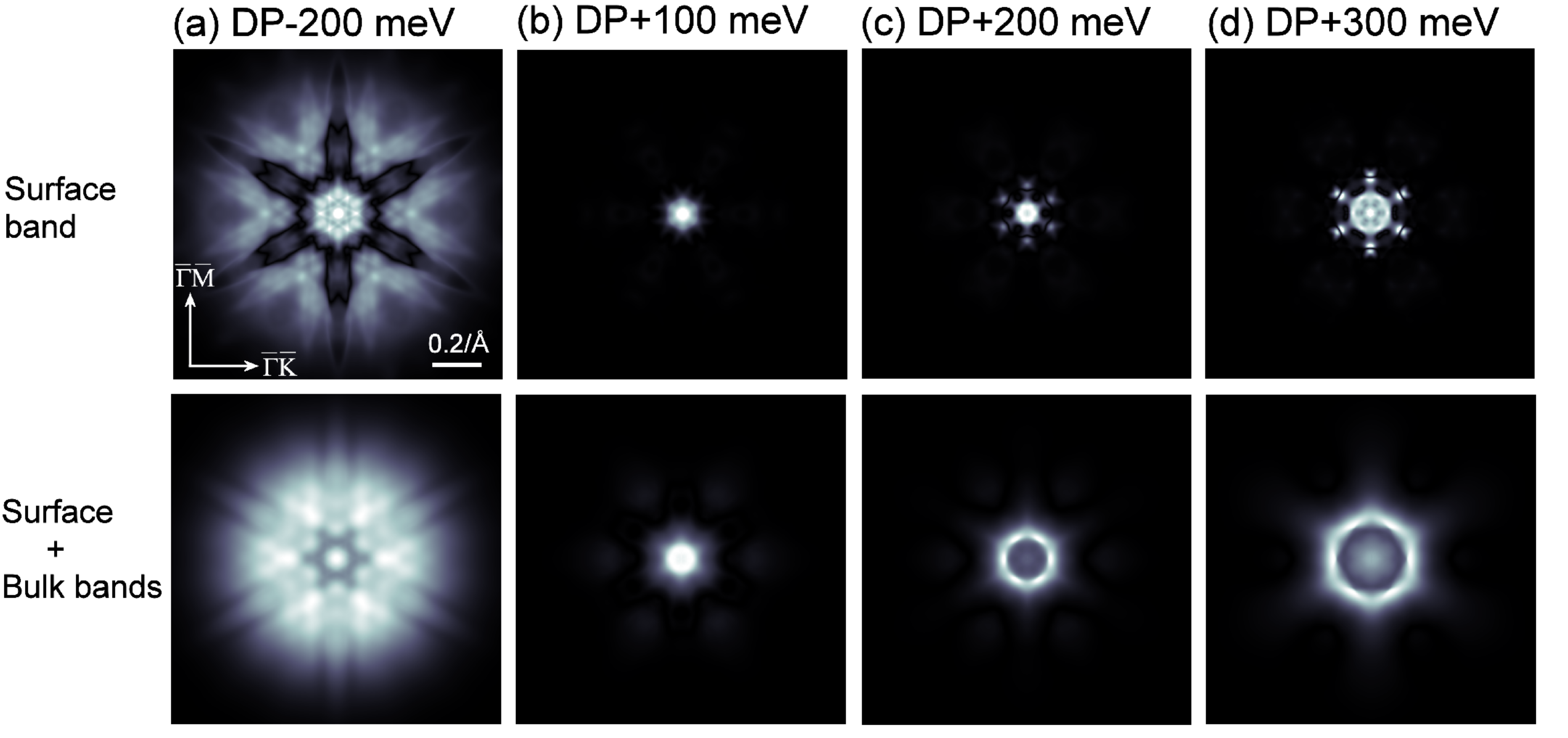} 
		\caption{Comparison of QPI patterns simulated using only the surface band (upper panel) and using the surface and all bulk bands (lower panel) at the energy denoted in the figure.}
		\label{fig:S7}
	\end{center}
\end{figure}

The QPI pattern, presented in Fig. 3(a), that corresponds to a 200 meV below the DP, was generated using a combination of surface and bulk states. In contrast, the remaining simulated patterns [Figs. 3(b)-3(d)] were derived solely from the surface state. Figure \ref{fig:S7} displays the QPI patterns simulated in two cases: one considering only the surface band, and the other considering both the surface and bulk bands. In the case of DP $-$200 meV, incorporating the bulk bands led to a significantly improved replication of the QPI pattern, closely aligning with the experimental data, in comparison to the inclusion of solely the surface state. [Fig. \ref{fig:S7}(a))]. Conversely, the incorporation of the bulk bands failed to reproduce QPI patterns that match the experimental ones above the DP [Figs. \ref{fig:S7}(b)-\ref{fig:S7}(d)]. This test outcome suggests that the Dirac cone exhibits a stronger tendency to hybridize with the bulk valence bands than with the bulk conduction bands.

\clearpage

\section{Energy dispersion of the surface state} 
\begin{figure}[h]
	\begin{center}
		\includegraphics[width=8.5cm]{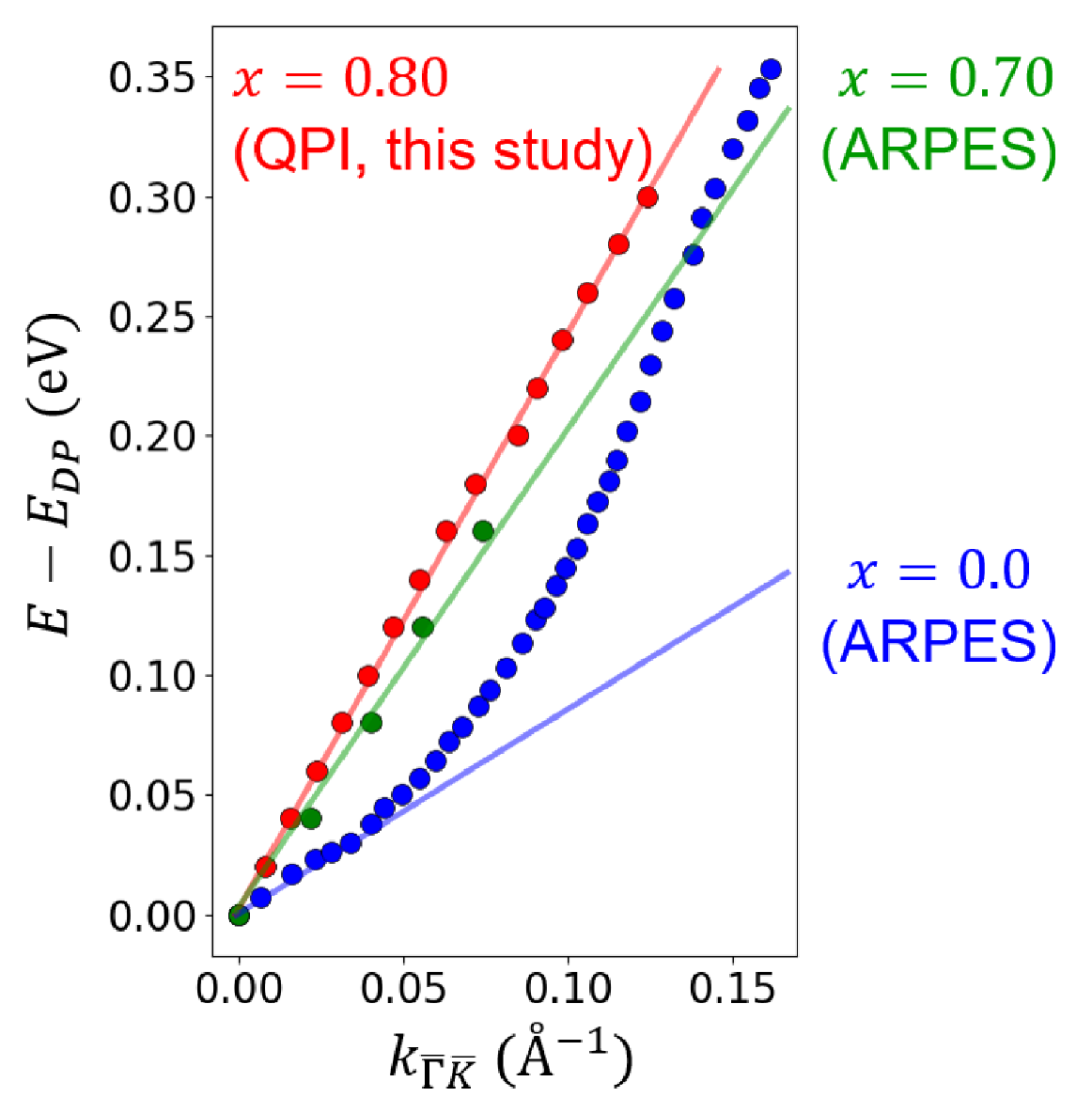} 
		\caption{Energy dispersions of the surface state in Pb(Bi$_{1-x}$Sb$_{x}$)$_{2}$Te$_{4}$ for different $x$ values. The band for $x$ = 0.80 was established from the QPI dispersion represented in Fig. 2(i). ARPES data for $x$ = 0.70 were acquired through our measurement \cite{Yaji2023} and that for $x$ = 0.0 was extracted from Fig. 2 in \cite{Kuroda2012}.}
		\label{fig:S8}
	\end{center}
\end{figure}

To extract the energy dispersion of the surface state from the QPI measurement results, it is crucial to identify the dominant scattering process. Based on the discussion in the main text and referring to Fig. 4, we have established the direction parallel to  the $\overline{\Gamma} \, \overline{M}$ direction (between one segment of the Dirac cone centered at the $\overline{\Gamma} \, \overline{K}$ direction and its next-nearest-neighbors) serves as the dominant scattering direction for Pb(Bi, Sb)$_{2}$Te$_{4}$. Assuming the validity of this scattering property is the vicinity of the Dirac point, we obtain the relation $q$ = $\sqrt{3} k_{\overline{\Gamma} \, \overline{K}}$, where $q$ represents the scattering vector of QPI and $k_{\overline{\Gamma} \, \overline{K}}$ denotes the momentum of the band along the ${\overline{\Gamma} \, \overline{K}}$ direction. The resulting band for Pb(Bi$_{1-x}$Sb$_{x}$)$_{2}$Te$_{4}$ ($x$ = 0.80) is depicted in Fig. \ref{fig:S8}, accompanied by ARPES data acquired through our own measurements ($x$ = 0.70) \cite{Yaji2023}, as well as referenced from Fig. 2 of \cite{Kuroda2012} ($x$ = 0.0). The band can be characterized by an almost linear dispersion for Bi-doped PbSb$_2$Te$_4$ ($x$ = 0.70 and 0.80), whereas the shape for PbBi$_2$Te$_4$ ($x$ = 0.0) appears to be non-linear. From the slope of the band, the group velocities at the Dirac point were estimated to be 2.2 and 1.0 eV\AA\ for Pb(Bi$_{0.20}$Sb$_{0.80}$)$_{2}$Te$_{4}$ and PbBi$_2$Te$_4$, respectively. These values reasonably agree with theoretical values derived from the DFT calculations [Fig. \ref{fig:S2}], which are 2.5 and 1.1 eV\AA\ for PbSb$_2$Te$_4$ and PbBi$_2$Te$_4$, respectively.

